\documentclass[modern]{aastex63}
\usepackage{amsmath}
\usepackage{diagbox}
\usepackage{soul}
\submitjournal{ApJ}

\shorttitle{NEI modeling in flares}
\shortauthors{Shen et al.}
\graphicspath{{./}{figures/}}
\begin{document}

%\title{Non-equilibrium Ionization Modeling of Petschek-type Shocks in Reconnecting Sheets in Solar Eruptions}
\title{Non-equilibrium Ionization Modeling of Petschek-type Shocks in Reconnecting Current Sheets in Solar Eruptions}

\correspondingauthor{Chengcai Shen}
\email{chengcaishen@cfa.harvard.edu}

\author[0000-0002-9258-4490]{Chengcai Shen}
\affiliation{Center for Astrophysics \textbar\ Harvard \& Smithsonian \\
60 Garden Street \\
Cambridge, MA 02138, USA}

\author[0000-0002-7868-1622]{John C. Raymond}
\affiliation{Center for Astrophysics \textbar\ Harvard \& Smithsonian \\
60 Garden Street \\
Cambridge, MA 02138, USA}

\author[0000-0001-6628-8033]{Nicholas A. Murphy}
\affiliation{Center for Astrophysics \textbar\ Harvard \& Smithsonian \\
60 Garden Street \\
Cambridge, MA 02138, USA}



\begin{abstract}

Non-equilibrium ionization (NEI) is essentially required for astrophysical plasma diagnostics once the plasma status departs from ionization equilibrium assumptions. 
In this work, we perform fast NEI calculations combined with magnetohydrodynamic (MHD) simulations and analyze the ionization properties of a Petschek-type magnetic reconnection current sheet during solar eruptions.
Our simulation reveals Petschek-type slow-mode shocks in the classical Spitzer thermal conduction models and conduction flux-limitation situations. The results show that under-ionized features can be commonly found in shocked reconnection outflows and thermal halo regions outside the shocks.
The departure from equilibrium ionization strongly depends on plasma density. In addition, this departure is sensitive to the observable target temperature: the high-temperature iron ions are strongly affected by NEI effects.
The under-ionization also affects the synthetic SDO/AIA intensities, which indicates that the reconstructed hot reconnection current sheet structure may be significantly under-estimated either for temperature or apparent width.
We also perform the MHD-NEI analysis on the reconnection current sheet in the classical solar flare geometry. Finally, we show the potential reversal between the under-ionized and over-ionized state at the lower tip of reconnection current sheets where the downward outflow collides with closed magnetic loops, which can strongly affect multiple SDO/AIA band ratios along the reconnection current sheet.

\end{abstract}



\section{Introduction} \label{sec:intro}
Magnetic reconnection, or the breaking and rejoining of magnetic field lines in a highly conducting plasma, is commonly believed to be a fundamental process during solar eruptions, and it plays a key role in rapid magnetic energy release \citep[e.g.,][]{Shibata2011LRSP....8....6S}.
It is commonly believed that reconnection involves rapid changes of temperature and density in reconnection regions of order $10^7$ K in solar coronal environments.
As an important observable signal of magnetic reconnection, such high-temperature reconnection current sheets have been frequently reported in literature with different instruments: SOHO/UVCS \citep[e.g.,][]{Ciaravella2002ApJ...575.1116C,ko2003,ciaravella2008}, SDO/AIA \citep[e.g.,][]{savage2010,reeves2011,Cheng2018ApJ...866...64C}, Hinode/EIS \citep[e.g.,][]{Warren2018ApJ...854..122W, Imada2021ApJ...914L..28I}.
Although the high-temperature reconnection outflow is predicted in both theoretical reconnection models and numerical experiments, the observable emission of the hot plasma (e.g., the emission from high ionization Fe lines) does not exactly reveal the temperature and density of the reconnection current sheet both in space and time due to non-equilibrium ionization (NEI) effects \citep[e.g.,][]{shen2013a}.
It is worth mentioning that the observable reconnection current sheets are due to hot plasma material emissions in reconnection regions, which may differ from the theoretical current sheet in models.
Therefore, we follow the term `current sheet' but refer to it as general dissipation and exhaust regions where the reconnection outflows and other relevant phenomena (e.g., slow-mode and fast-mode shocks) are expected in different plasma environments.

Around the reconnection current sheet, if plasma is rapidly heated due to reconnection, then the ion charge states would correspond to lower temperatures than the actual temperature for a sizeable duration longer than the dynamic time-scale, and the plasma is referred as under-ionized. If plasma cools rapidly (e.g., due to adiabatic expansion), then the ionization states would correspond to higher than the actual temperature, and the plasma is over-ionized.
In either case, if the plasma is in an NEI state, its ionization state must be obtained by considering time-dependent ionization. 
For instance, in a typical high-temperature solar flare environments with temperature $T \sim 10^7$ K, plasma density $n_e \sim 10^9$ cm$^{-3}$, magnetic field strength $B \sim 50$ G, and characteristic length $L \sim 50$\ Mm, the Alfv\'{e}nic transit time is about 15 s, and the thermal conduction time-scale is $\sim$ 25 s. The above two typical dynamic time-scales are generally much shorter than the ionization (and recombination) times ($\sim 10^3$s) according to the characteristic equilibration density-weighted time-scale, of order $10^{12}$ cm$^{-3}$s, for Fe ions \citep{Smith2010ApJ...718..583S}. 

NEI effects have been reported in the post coronal mass ejection (CME) plasma during solar eruptions in several models.
\cite{ko2010} modeled the UV and X-ray emission including non-equilibrium ionization states in a Petschek-type reconnection flow in the post-CME current sheet. They found that the predicted emission intensities are consistent with SOHO/UVCS and Hinode/X-ray observations.
\cite{shen2013a} performed a post-processed NEI analysis in a large scale reconnection current sheet beneath the erupted CME in an MHD model. They calculated the time-dependent ionization states using the numerical simulation data from \cite{Reeves_2010ApJ...721.1547R}, and predicted the UV/EUV emission intensity. It was found that ionization charge states can be significantly under-estimated (or over-estimated) by up to a factor of two compared with the assumption of equilibrium ionization (EI) along the current sheet. 
Recently, \cite{Lee2019ApJ...879..111L} investigated the hot plasma sheet observed by SDO/AIA and Hinode/XRT on 2012 January 27. They calculated intensity ratios between different narrow-band images and compared them with the NEI modeling emission, and suggested that the ratio-ratio plot for intensities in different passbands can be used to express how far the plasma departs from the ionization equilibrium.
For a large solar flare on 2017 September 10 observed by Hinode/EIS, \cite{Imada2021ApJ...914L..28I} reported that observed \ion{Fe}{24}/\ion{Fe}{23} ratios are consistent with time-dependent ionization effects at a constant electron temperature ($\sim$25 MK). This research also pointed out that NEI is required to investigate the temperature gradient in the magnetic reconnection region from density, temperature, and velocity diagnostics.

In general, the plasma inside the reconnection region is expected to be in an under-ionized state due to rapid heating processes that occur during reconnection. 
For example, \cite{Imada_2011ApJ...742...70I} analyzed the time-dependent ionization in steady Petschek-type reconnection models. They pointed out that iron ions are mostly in non-equilibrium ionization states in the reconnection region, and line emissions are significantly different from those determined from the equilibrium ionization assumption.
%However, there are several important unsolved questions, including: 
However, an important question is still not well understood:
how does NEI affect the fine structure of observable current sheets in various realistic magnetic reconnection regions? 
In recent decades, a set of reconnection models have been proposed, but which model drives solar eruptions is still an open question. 
Petschek-type reconnection \citep{Petschek1964NASSP..50..425P, Forbes1987RvGeo..25.1583F} is an important mechanism in solar atmosphere among different reconnection models, such as the classical Sweet-Parker current sheet \citep{Sweet_1958IAUS....6..123S, Parker1957JGR....62..509P} and plasmoid instabilities that develop inside it \cite[e.g.,][]{Loureiro_2007PhPl...14j0703L, Ni_2010PhPl...17e2109N, Bhattacharjee_2009PhPl...16k2102B, huang2016}, and turbulent reconnection 
where the magnetic energy is supposed to be cascading in small-scale structures and heats the diffusion region \citep[e.g.,][]{Lazarian1999ApJ...517..700L}.
Recently, \cite{Lin2021PhPl...28g2109L} pointed out that Petschek-type magnetic reconnection can be induced using a simple resistivity gradient in the reconnection outflow direction.
The localized X-point is also predicted by \cite{Forbes_2018ApJ...858...70F} in the theoretical analysis of large-scale CME/flare eruptions.
Therefore, studies of the Petschek-type reconnection and detailed emission features in NEI are important to investigate plasma properties of the reconnection current sheet.
%[Mention turbulent reconnection model of Lazarian and Vishniac 1999? How does heating on very smallsscales differ from large scale shocks?]

The high temperatures inside the reconnection current sheet can be produced by several processes including Ohmic heating in reconnection sites (normally referred as X-points with the highest current), adiabatic compression, thermal conduction, slow-mode shocks along the two sides of the current sheet, and fast-mode shocks where super-magnetosonic reconnection outflows impact closed magnetic loop structures such as flare loops. In order to accurately simulate the ionization states and predict the corresponding emission of a reconnection current sheet, it is important to incorporate both the above physical processes and time-dependent ionization in a self-consistent model.
In our previous models \citep{shen2013b,Shen2015,Shen2017ApJ...850...26S}, we have employed post-processed methods to perform NEI calculations using MHD simulation results, especially temperature, density, and plasma flow evolution history. However, the post-processed NEI method depends on the streamline trajectory in which errors could be introduced due to the flow field interpolation in time (and space). It is also difficult to trace plasma evolution in turbulent flow regions, and it is not possible to self-consistently compute the radiative cooling rate during the MHD simulation.
Therefore, an NEI calculation that can be incorporated into the MHD model itself is required for performing accurate NEI analysis in the reconnection current sheet. This kind of ionization solver has been reported in other astrophysical research regimes. \cite{Orlando2003MmSAI..74..643O} investigated the non-equilibrium ionization effects in the compact flare loop by combining the NEI calculation within hydrodynamic (HD) modeling. An NEI solver \citep{Zhang_2018ApJ...864...79Z} using the eigenvalue method \citep[e.g.,][]{masai1984,hugheshelfand1985,Smith2010ApJ...718..583S} and the AtomDB database \citep{Foster_atomdb_2018} has been developed to perform fast NEI calculations in the FLASH code \citep{Fryxell2000ApJS..131..273F}, and has been used to analyze the non-equilibrium ionization in supernova remnants \citep[e.g.,][]{Zhang2019ApJ...875...81Z}. Charge states in the solar wind have been calculated using the plasma electron temperature, density, and velocity in several solar wind simulations such as the Michigan Ionization Code (MIC) model \citep[e.g.,][]{Landi2012ApJ...744..100L,Landi2015ApJ...812L..28L} and Magnetohydrodynamic Algorithm outside a Sphere (MAS) model \citep[e.g.,][]{Lionello2019SoPh..294...13L}.

In this paper, we describe a model that combines the non-equilibrium ionization (NEI) calculation in MHD simulations in a public code: Athena \citep{Stone2008ApJS..178..137S}. In Section \ref{sec:method}, we briefly summarize the NEI solver in the Euler framework and the MHD modeling setup. In Section 3, we analyze the NEI properties around the Petschek-type reconnection current sheet, and predict EUV emissions. Then we discuss how the asymmetrical configuration of the reconnection current sheet affects the NEI properties. The discussion and conclusions are given in Sections 4 and 5.

\section{Description of the Numerical Method} \label{sec:method}
%%\subsection{Time-dependent Ionization}
In this section, we describe how to include the NEI solver in a well-developed MHD code. In this paper, we focus on the application using the Athena code, but this process will be similar to other grid-based MHD codes.
The governing resistive MHD equations in our combined MHD-NEI simulations are as the following:
\begin{eqnarray}
    \frac{\partial\rho}{\partial t} & + & \nabla\cdot(\rho \mathbf{v})=0, \label{eq:mass} \\
    \frac{\partial\rho\mathbf{v}}{\partial t} & + & \nabla\cdot(\rho\mathbf{v}\mathbf{v}-\mathbf{B}\mathbf{B}+\mathbf{P^{*}})=0, \label{eq:momentum} \\
    \frac{\partial \mathbf{B}}{\partial t} & - & \nabla\times( \mathbf{v}\times \mathbf{B})=\eta_{m}\nabla^{2} \mathbf{B}, \label{eq:induction} \\
    \frac{\partial E}{\partial t} & + & \nabla\cdot[(E+P^{*})\mathbf{v} - \mathbf{B}(\mathbf{B}\cdot\mathbf{v})]=S, \label{eq:energy}
\end{eqnarray}
where ${\mathbf{P^{*}}}$ is a diagonal tensor with components $\mathbf{P^{*}} = P + B^{2}/2$ (with $P$ the gas pressure), and $E=\frac{P}{\gamma-1} + \frac{1}{2}\rho v^2 + \frac{B^2}{2} $ is the total energy density, ${\gamma = 5/3}$ is the adiabatic index, and the energy source term $S = \mu_0 \eta_{m}{j^2} + \nabla_{\|}\cdot\kappa\nabla_{\|} T$, which includes Ohmic dissipation and thermal conduction.
The quantities $\rho$, $\mathbf{v}$, $\mathbf{B}$, and T are mass density, flow velocity, magnetic field, and temperature, respectively.
Here $\mu_0, \eta_m$, and $\kappa$ are the magnetic permeability of free space, magnetic diffusivity, and the parallel component of the Spitzer thermal conduction tensor.

The time-dependent ionization equations in an Eulerian framework can be described as the following:
\begin{equation}
    \frac{\partial f_{i}}{\partial t} + \nabla\cdot f_{i} \mathbf{v} = n_e \left[C_{i-1}f_{i-1} - \left(C_i+R_i\right)f_i +R_{i+1}f_{i+1}\right],
\label{eq:nei}
\end{equation}
where $f_i$ is the ion fraction of the $i_{th}$ ionization state, $C_i$ and $R_i$ are ionization and recombination rate coefficients for these ions. $n_e$ is the electron density, which is $\sim 1.2$ proton density for fully ionized plasma in our single fluid MHD model. For an element with the atomic number $Z$, $f_i$ covers all $Z+1$ ion charge states. Here, the $C_i$ and $R_i$ rates are functions of temperature which is computed from the MHD simulations at each time-step. 
In optically-thin plasma, such as solar corona, the most important ionization/recombination processes are collisional ionization, excitation-autoionization, radiative recombination, and dielectronic recombination. Therefore, we consider the above electron temperature (and density) dependent ionization and recombination rates in our models.
Because the right side of Equation (\ref{eq:nei}) does not explicitly involve the time (and spatial) operations, we apply the operator splitting method and set it as an NEI source term. In this way, Equation (\ref{eq:nei}) can be separately solved in two steps: advection part ($\frac{\partial f_{i}}{\partial t} + \nabla\cdot f_{i} \mathbf{v} = 0$), and source part (${\frac{\partial f_{i}}{\partial t} = n_e \left[C_{i-1}f_{i-1} - \left(C_i+R_i\right)f_i +R_{i+1}f_{i+1}\right]}$). For the advection part, it is easy to apply a similar scheme in the MHD code itself. In fact, it is even more convenient to solve the advection part through the passive scalars in the MHD mass equation (Equation  (\ref{eq:mass})), which is usually introduced in most MHD codes. In the Athena code, the mass equation with passive scalars is:
\begin{equation}
    \frac{\partial (s_i \rho)}{\partial t} + \nabla\cdot(s_i \rho \mathbf{v})=0, 
\end{equation}
where $s_i (i=1,..., N)$ are the mass fractions of $N$ passive scalars. The ion fraction for a single particular element with $N+1$ charge states can be directly obtained as: $f_i = {s_i}/{\sum{s_i}}$.

\subsection{Eigenvalue method}
We solve the source part of the time-dependent ionization equations using the eigenvalue method \citep[e.g.,][]{hugheshelfand1985,Smith2010ApJ...718..583S, Shen2015}, in which ionization equations can be represented in the matrix form, and the exact exponential solution can be calculated using matrix multiplication. 
The time-dependent ionization equation can then be written as:
\begin{equation}
    \frac{\partial \mathbf{F}}{\partial t} = n_e \mathbf{A} \cdot \mathbf{F},
    \label{eq:nei_matrix}
\end{equation}
where $\mathbf{F}$ is a vector containing ion fractions $f_i$, and $\mathbf{A}$ is the matrix containing ionization and recombination rate coefficients ($C_i$ and $R_i$) on the right side of Equation (\ref{eq:nei}). Using the eigenvalues and eigenvectors of the matrix $\mathbf{A}$, the solution of Equation (\ref{eq:nei_matrix}) can be presented in the form:
\begin{equation}
    \frac{\partial \mathbf{F'}}{\partial t} = n_e \lambda \cdot \mathbf{F'},
    \label{eq:nei_eigen}
\end{equation} 
where $\lambda$ is a diagonal matrix containing eigenvalues of the matrix $\mathbf{A}$. By defining $\mathbf{V}$ as the matrix of eigenvectors of $\mathbf{A}$ with all eigenvalues, $\mathbf{F'}$ can then be defined as $\mathbf{V}^{-1} \cdot \mathbf{F}$.
For a particular temperature ($T_c$), it is easy to get corresponding eigenvalues and eigenvectors. The solution of Equation (\ref{eq:nei_eigen}) can be reduced to simple exponentiation: $\mathbf{F'} = \mathbf{F_0} \exp(-{n_e}\lambda(T_c) t)$. Here $\mathbf{F}_0$ is the initial condition containing ion fractions for this element. 
Hence, the ion fraction vector $\mathbf{F}$ for the ionization state at any time is easy to compute by the matrix multiplication using $\mathbf{F'}$ and the eigenvector matrix at temperature ${T_c}$. This method is very robust and has the advantage over explicit methods that take a very long time-step at a single temperature would result in an equilibrium ionization state. 
Once we pre-calculated eigenvalues and eigenvectors for all ions on a temperature grid, the code can reload eigenvalues and eigenvectors, and quickly perform the above calculation.

We follow this eigenvalue method and apply our well-tested NEI code \citep{Shen2015} in the following combined MHD-NEI simulations. This NEI code\footnote{NEI code: \url{https://github.com/ionizationcalc/time_dependent_fortran}} was originally developed using Fortran and has been restructured into a C module that can efficiently perform the in-line NEI calculation in Athena. We also use updated atomic data from the Chianti database \citep[Version 9,][]{Dere_2019ApJS..241...22D}
%[need reference for version number]
to create lookup tables, including all necessary ionization and recombination rates, and eigenvalue corresponding matrices.
We employ this method in several test projects, such as one-dimensional shock tubes, and compare the results with the post-processed NEI calculations (see Appendix A for details). The result shows that this in-line NEI module calculates charge states accurately, and it can be used in complex problems such as shocks.

\subsection{MHD Model Setup}
% initial parallel current sheet
% perturbation forms
% localized resistivity, formula
% with Bz? and uniform background density

% initial condition: 
% perturbation 
% uniform resistivity
% without Bz. High pressure inside the CS

We set up two types of MHD models to study reconnection outflows: (i) steady Petschek-type reconnection, and (ii) realistic reconnection current sheet during solar flares.
In model (i), the initial condition consists of a pre-existing Harris-type current sheet along the y-direction with the non-dimensional width $w=0.025$ as follows:
\begin{eqnarray}
    B_x(x, y) = 0, \\
    B_y(x, y) = \tanh(\frac{x}{w}), \\
    B_z(x, y) = (1-B_y(x, y)^2)^{1/2}, \\
    p(x, y) = {\beta}_0/2, \\
    {\rho}(x, y) = 1.
\end{eqnarray}
Here $\beta_0$ is the background plasma $\beta$ (the ratio of the magnetic pressure to gas pressure) in the ambient region. The initial current sheet is then in dynamical equilibrium with uniform temperature and density. 
In order to have the system evolve rapidly from the initial steady state to a fast reconnection phase, we introduce a perturbation magnetic field $B_{1x}$ and $B_{1y}$ on this preexisting current sheet as follows:
\begin{eqnarray}
    B_{1x}(x, y) = \frac{2\pi}{L_y}A_{pert}\cos(\frac{\pi x}{L_x})\sin(\frac{2\pi (y-y_c)}{L_y}) B_0, \\
    B_{1y}(x, y) =  -\frac{2\pi}{L_x}A_{pert}\sin(\frac{\pi x}{L_x})\cos(\frac{2\pi (y-y_c)}{L_y}) B_0.
\end{eqnarray}
Here $A_{pert} = 0.0001$ is the non-dimensional perturbation strength located at $[x=0.0, y_c=0.0]$. $L_x$ and $L_y$ are non-dimensional perturbation wavelengths that are set to 2 and 2 in order to minimize perturbations at boundaries. We set the symmetrical boundary condition along the center of the system ($x=0$) and the open boundary condition at other sides. The simulation domain then covers the right-half of the reconnection region, ranging from $x=0$ to $1$ and $y=-1$ to $1$. To drive Petschek-type reconnection, we introduce an anomalous strong diffusion at the system center with double-Gaussian distribution in space where the equivalent magnetic Reynolds number is about $10^4$. Driven by the initial perturbation on magnetic fields and the localized enhanced diffusion at the system center, the initial current sheet gradually develops into the single X-point reconnection geometry associating with a pair of slow-mode shocks extended along current sheet edges. 

Our second type of reconnection model is based on the classical two-ribbon flare configuration, where the magnetic reconnection is expected to appear in a vertically extended current sheet above the reconnected post-flare loops, which is also referred to as the classical CSHKP model\citep{Carmichael1964, Sturrock1968, Hirayama1974, Kopp-Pneuman1976}. Following our previous modeling setups \cite[e.g.,][]{shen2011, Shen_2018ApJ...869..116S}, we also start the simulation from a thermal and dynamical equilibrium current sheet structure but include the line-tied boundary condition at the lower side to represent the solar surface where the magnetic field is rooted into the boundary and the plasma cannot slip. To simulate Petschek-type shocks, we also apply a locally enhanced resistivity as in the above discussion.

%% Thermal conduction: general setting, flux limitation-> results, saturation limitation around the shock
\subsection{Thermal Conduction}
Thermal conduction can significantly affect plasma properties during magnetic reconnection. In numerical simulations, one of the direct impacts is that the conduction along magnetic field lines may cause wider reconnection outflows \citep{Yokoyama1997ApJ...474L..61Y}.
A high-temperature plasma region (also referenced as thermal halo) has been proposed in both analytical and numerical models \citep[e.g.,][]{Seaton2009ApJ...701..348S, Yokoyama2001ApJ...549.1160Y}.
However, the thermal conductive flux could be overestimated in the shock front where the temperature jump is extremely sharp and the corresponding conductive flux based on the Spitzer coefficients becomes nonphysical. There is also a potential jump at a collisionless shock that inhibits electron transport from downstream to upstream.
In general, the conductive flux along the magnetic field lines is given by $-\nabla_{\|}\cdot\kappa\nabla_{\|} T$, where the Spitzer conductivity $\kappa$ for elections is the non-linear function of plasma temperature in the form of 
\begin{equation}
    \kappa = 10^{-6}T^{5/2} ~\rm erg~K^{-1}~cm^{-1}~s^{-1}.
\end{equation}
The above classical coefficient is valid under the assumption that the mean free path ($L_{mfp}$) of electrons is much shorter than the temperature scale length ($L_T = T/|\nabla T|$). For extremely sharp temperature changes such as a shock front, the mean free path becomes comparable to or larger than the temperature scale length. 
Furthermore, non-thermal particles are expected to be accelerated in shocks, so accurately resolving the conductive flux requires detailed information on particle velocity distributions around the shock fronts. 
Instead of discussing the thermal conductive flux in various non-Maxwellian particle velocity distributions, we focus on the ionization behavior inside the high-temperature reconnection current sheet in this work. 
Therefore, we introduce an additional conductive flux limitation on the classical Spitzer flux form to limit the nonphysical thermal conduction around the shock front.
We set the upper limit of conductive flux to be the saturation flux \citep[e.g.,][]{cowie_evaporation_1977} when the electron mean free path is sufficiently large ($L_{mfp} > L_{T}$).
Following the method reported by \cite{winter_simulating_2011} in one-dimensional hydrodynamic models, we also introduce a power-law factor in the Spitzer conductive flux when the mean free path is close to the temperature scale length \citep{Rosner1986psun....2..135R}:
\begin{equation}
    F_{cond} \sim  0.11 \frac{L_{mfp}^{-0.36}}{L_{T}} F_{Spizter},
    \label{eq: cond86}
\end{equation}
where $F_{Spizter}$ is the Spitzer conductive flux and $F_{cond}$ is the modified flux.
This approach allows us to obtain more realistic temperature distributions inside the reconnection current sheet, which is crucial for the NEI analysis.
However, the conduction may depart more from the above Equation (\ref{eq: cond86}) due to the micro-instabilities in the above flux limitation models \citep{Rosner1986psun....2..135R}, which is outside the scope of this paper, and will be investigated in future work.
For comparison,  we also run the same model but with the classical Spitzer thermal conduction, which gives insight into how the thermal halo can develop in such an extremely strong thermal conduction situation.
Therefore, the conductive flux-limited simulation can be thought of as the lower-limit of the conduction effects, while the classical Spitzer heating flux model is close to the upper-limit of the thermal conduction modeling.

\begin{table}[]
\centering
\caption{Primary Simulation Parameters for Different Cases} 
\label{tab:parameters}

\begin{tabular}{|c|c|c|c|}
\hline
\diagbox[innerwidth=6cm, innerleftsep=-5pt]{Plasma $\beta$}{Density} & {$5 \times 10^9$ cm$^{-3}$} & {$1 \times 10^9$ cm$^{-3}$} & {$5 \times 10^8$ cm$^{-3}$} \\
\hline
{0.04} & Case A1                          & Case B1, D       & Case C1 \\
\hline
{0.05} & Case A2, A2$^*$         & Case B2        & Case C2 \\
\hline
{0.075} & Case A3                             & Case B3        & Case C3 \\   
\hline
{0.1} & Case A4                             & Case B4       & Case C4 \\
\hline
\end{tabular}
\\ Notes: Case A2$^*$ used the Spitzer conduction coefficient. Case D includes gravity in the $y-$ direction with plasma $\beta = 0.04$ and density $n_e = 1.0\times 10^9$ cm$^{-3}$ at $y=1.0L_0$. \\
\end{table}

\section{Results} \label{sec:results}
The primary simulation parameters for different cases are listed in Table \ref{tab:parameters}.
Figure \ref{fig:13gh_halo_vs_nohalo} shows the Petschek-type magnetic reconnection configuration in our simulations. After the initial perturbation and the early slowly magnetic diffusion process, the reconnection process is close to the steady phase at the time $t \sim 6t_0$ (for example, $\sim$ 520 s in Case A2 with the chosen characteristic time $t_0 = 87$\ s), when the reconnection inflow and outflow speed does not significantly change.
The reconnection X-point is located at the origin ($x=0, y=0$) due to the localized magnetic resistivity at this position. From the X-point to two sides of the current sheet, a pair of slow-mode shocks can be clearly identified where the magnetic field dramatically changes and the plasma is abruptly heated to more than $\sim 10^7$K. These shock fronts also can be seen from plasma profiles crossing the edge of the current sheet, and the shock jump conditions are met with the finite shock thickness in 2 $\sim$ 3 MHD simulation grid cells (see more details in Appendix B and Figure \ref{fig:sk_thermalhalo}). At this time, the reconnection inflow speed ($\sim V_x$) is about 100 km/s while the outflow velocity ($V_y$) can reach the Alfv\'{e}n speed of $>1000$km/s. This configuration therefore shows a quick reconnection process with the non-dimensional reconnection rate $M_A \sim$ 0.1 ($M_A \equiv V_{inflow}/V_{outflow}$).

As shown in Table \ref{tab:parameters}, we ran a set of combined MHD-NEI simulations for different densities and temperatures.
We chose typical solar active region parameters, with the initial background temperature of $2 \times 10^6$ K in all cases.
The temperature inside the reconnection current sheet then depends on $\beta$, and can be estimated according to the shock jump condition:
\begin{equation}
    \frac{T_{Dn}}{T_{Up}} = 1 + \frac{2}{5\beta}.
\end{equation}
Here $T_{Dn}$ is the downstream (or shocked plasma) temperature inside the reconnection current sheet, and $T_{Up}$ is the upstream (or the background coronal) temperature.
%For the high temperature solar coronal environments, 
We change the plasma $\beta$ from 0.04, 0.05, 0.075 to 0.1 to obtain postshock temperatures $2.2 \times 10^7, 1.8 \times 10^7, 1.267 \times 10^7,  10^7$ K, respectively.
Due to the magnetic energy dissipation and the reduction of the ambient field during the system evolution, the postshock temperatures measured in the following sections decline slightly to around $1.8 \times 10^7, 1.6 \times 10^7, 10^7,  8 \times 10^6$ K until the system evolves into a phase with relatively steady reconnection.
The plasma density can be expected to increase a factor of 2.36 $\sim$ 2.2 in shocked regions for the above $\beta$ list, according to the formula $\rho_{Dn}/\rho_{Up} = 5(1 + \beta)/(2+5\beta)$ \citep[e.g., in][]{ko2010}.
Because the ionization (or recombination) timescale is proportional to the plasma density, models with low density are expected to cause more strong NEI effects. We then set the ambient density to $5 \times 10^9$cm$^{-3}$ in Case A series for the typical coronal environment, but also run the more tenuous density of $10^9$ cm$^{-3}$ and $5 \times 10^8$ cm$^{-3}$ in Case B series and C series for comparison.

\subsection{Under-ionzied Plasma}
% case g and case h: show charge states crossing the shock front
\begin{figure}
    \centering
    \plotone{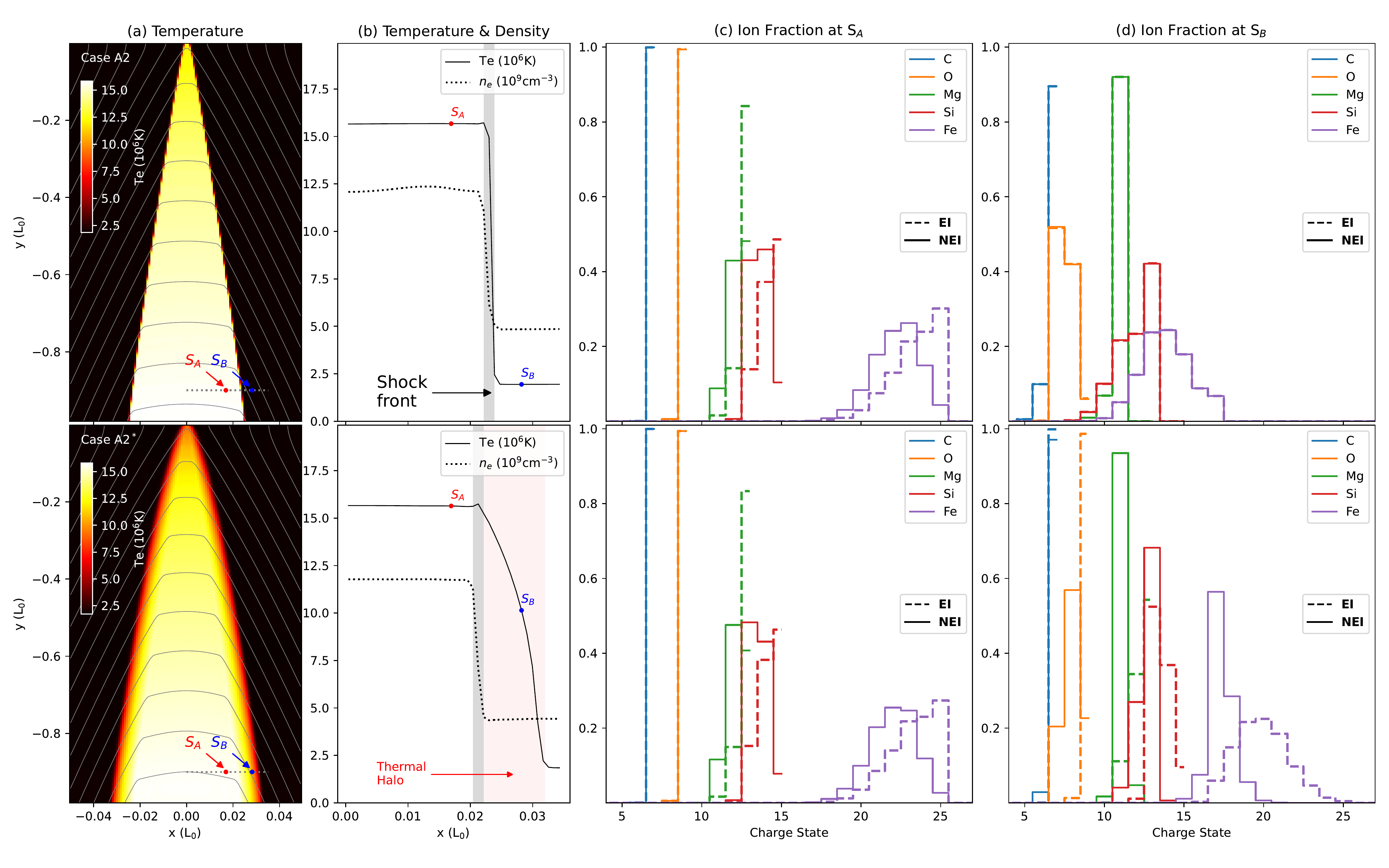}
    \caption{Temperature and density distribution across the Petschek-type shocks in Case A2 and A2$^*$. Here, Case A2 employed the conductive flux-limitation and Case A2$^*$ is for the classical Spritzer conduction simulation.  Two sampling points ($S_A$, $S_B$) have been chosen along the dotted horizontal sampling line at $y=-0.9L_0$. Here, the non-dimensional characteristic length $L_0 = 1.5 \times 10^8$m. Panels (b) are temperature and density profiles along the sampling line shown in (a). The gray and red shaded regions indicate the shock front and thermal halo regions, respectively. 
    Panels (c-d) show ion fractions of C, O, Mg, Si, and Fe at $S_A$ and $S_B$. The solid lines show the NEI results, and the dashed lines are for equilibrium ionization. The charge state in the horizontal axis is in spectroscopic notation in which  \ion{Fe}{12} means Fe $^{11+}$.}
    \label{fig:13gh_halo_vs_nohalo}
\end{figure}

In this section, we analyze the ionic charge state in the Petschek-type reconnection current sheet and surrounding regions, and show how the ionization state departs from ionization equilibrium. We first compare Case A2 and Case A2$^*$ with the conduction flux limit simulation and classical Spitzer conduction coefficients, respectively.
Magnetic configurations are similar among the two cases, as shown in Figure \ref{fig:13gh_halo_vs_nohalo}(a).
In the downstream region (or inside the current sheet), the plasma is heated to $\sim$15.7\ MK from the background coronal temperature $\sim 1.95$\ MK and $\sim 1.85$\ MK for Case A2 and A2$^*$, respectively. The plasma density jumps from $\sim 5 \times 10^9$ to ~$1.2 \times 10^{10}$\ cm$^{-3}$ crossing the shock front (Figure \ref{fig:13gh_halo_vs_nohalo}b). 
Outside the current sheet (or in the upstream region of the shocks), a temperature `halo' region (red shading in Figure \ref{fig:13gh_halo_vs_nohalo}(b) is clear in Case A2$^*$ due to the strong thermal conduction. On the other hand, once the thermal conduction flux-limitation is employed in Case A2, the thickness of the thermal halo decreases to the grid size. Case A2 also appears to have a slightly higher temperature/density compared with Case A2$^*$ inside the reconnecting current sheet due to the reduced thermal conduction.

We then investigate the charge state distribution at two typical sampling points ($S_A$ and $S_B$) in the shock downstream (inside reconnection current sheet) and upstream (in ambient regions), respectively. Figure \ref{fig:13gh_halo_vs_nohalo}(c-d) shows the ion populations for chosen abundant elements in the solar corona, including C, O, Mg, Si, and Fe.
Here, the solid lines are for time-dependent ionization results solved by the in-line NEI module, and dashed lines indicate ionization equilibrium which depends only on the local temperature.
At the point $S_A$, because the plasma is suddenly heated by the shock, ion populations of high charge states (e.g., \ion{Fe}{24}, \ion{Fe}{25}) are clearly lower than the equilibrium ionization. The departure of NEI ion fractions from ionization equilibrium can be found in the ionization distribution profiles, where solid lines (for NEI) are skewed toward lower charge states compared to equilibrium ionization results (dashed lines). This result indicates that the plasma is clearly under-ionized. 
In the ambient region ($S_B$) beyond the thermal halo, on the other hand, the profiles are the same between NEI and equilibrium ionization in Case A2 due to the unchanged plasma temperature and density.
In the thermal halo region for Case A2$^*$, the NEI profile at $S_B$ also appears under-ionized features because this plasma is pre-heated from the ambient corona temperature before it reaches the shock front. 

\begin{figure}
    \centering
    \includegraphics[width=0.5\textwidth]{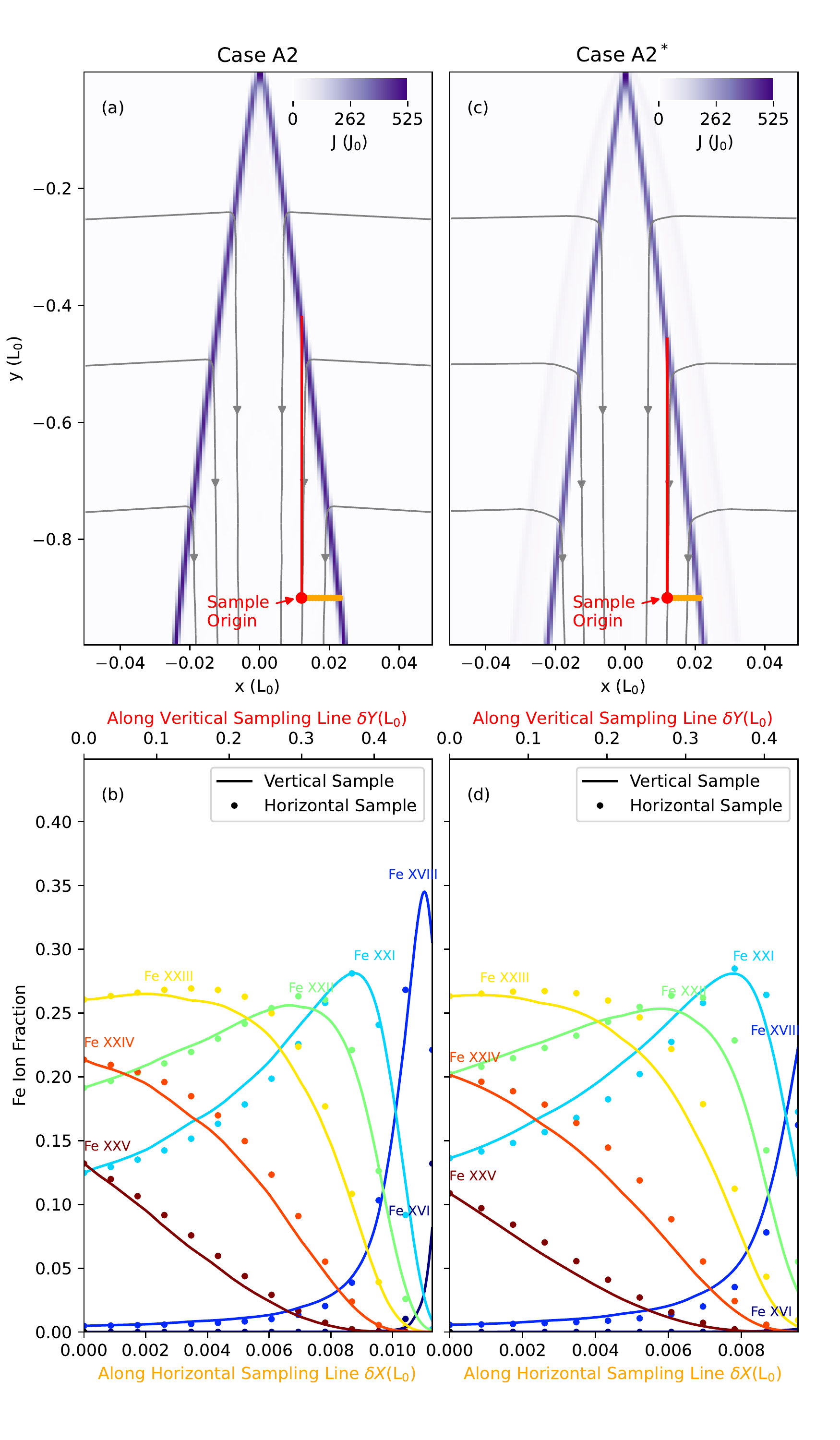}
    \caption{Iron ion fraction distribution across Petschek-type shocks in Case A2 and A2$^*$. In the top panels, the gray streamline shows plasma flows. The strong current density ($J$) indicates the position of shock fronts. In panels (b) and (d), ion fraction distributions inside reconnection current sheets are shown along the two chosen sampling lines in vertical and horizontal directions.
    }
    \label{fig:2}
\end{figure}

The two-dimensional spatial distributions of ion fractions inside the Petschek-type reconnection current sheet are shown in Figure \ref{fig:2}.
In this figure, the high current density indicates the edges of the shock front in Figure \ref{fig:2}(a)(c) and the gray lines with arrows are streamlines of velocity fields in MHD simulations for Case A2 and Case A2$^*$.
It is clear that the bulk plasma inflow towards the shock front is roughly along the horizontal direction ($x-$ direction) outside the current sheet, and then rapidly turns to the reconnection outflow direction ($y-$ direction) as it flows through the shock front. 
This behavior follows the Petschek reconnection theoretical expectation.
In an ideal Petschek reconnection current sheet, the postshock density, temperature, and velocity are uniform in both $x-$ and $y-$ directions, the ionic distribution in $x-$ direction, therefore, entirely depends on flow paths in $y-$ direction as shown by the solid red and orange color sampling lines in Figure \ref{fig:2} (a)(c). 
In other words, the ion's distribution along the red and orange sampling lines would be exactly the same in ideal situations.
We then compare the chosen Fe ion fraction along the above two sampling lines starting from a random origin point, and ending on the shock front (Figure \ref{fig:2}(b)(d)).
The overall feature is that these two distributions (the dot-lines and solid lines) basically match each other.
The distances of a sampling origin point away from the shock fronts can be decided by Petschek shock angle as 
\begin{equation}
    \Delta_y = \frac{\Delta_x}{tan(\theta)}.
\end{equation}
Here $\Delta_y$ and $\Delta_x$ are distances to the shock front in $y-$ and $x-$ directions, and $\theta$ is the shock angle in downstream (also see the intersection angle between $y-$axis and the shock edges in Figure \ref{fig:2}).

In general, the plasma conditions are not uniform inside the current sheet in realistic models due to compression, which causes a slightly higher temperature away from the X-point (Figure \ref{fig:13gh_halo_vs_nohalo}). The velocity streamlines also tend to bend toward the system center ($y-$axis) as shown in Figure \ref{fig:2}(a)(c).
Thus, parts of sampling points along the horizontal orange sampling line show larger fractions on high ionic states compared with the vertical sampling line.
This suggests that 2D distributions of ion populations inside the current sheet could be approximated by analyzing the plasma evolution along $y-$ directions in Petschek-type reconnection configurations \citep[e.g.,][]{Lee2019ApJ...879..111L}. However, combined MHD-NEI modeling is essentially required to obtain accurate ion distributions.

\subsection{Ionization Properties vs. Temperature and Density}

\begin{figure}
    \includegraphics[width=1.0\textwidth]{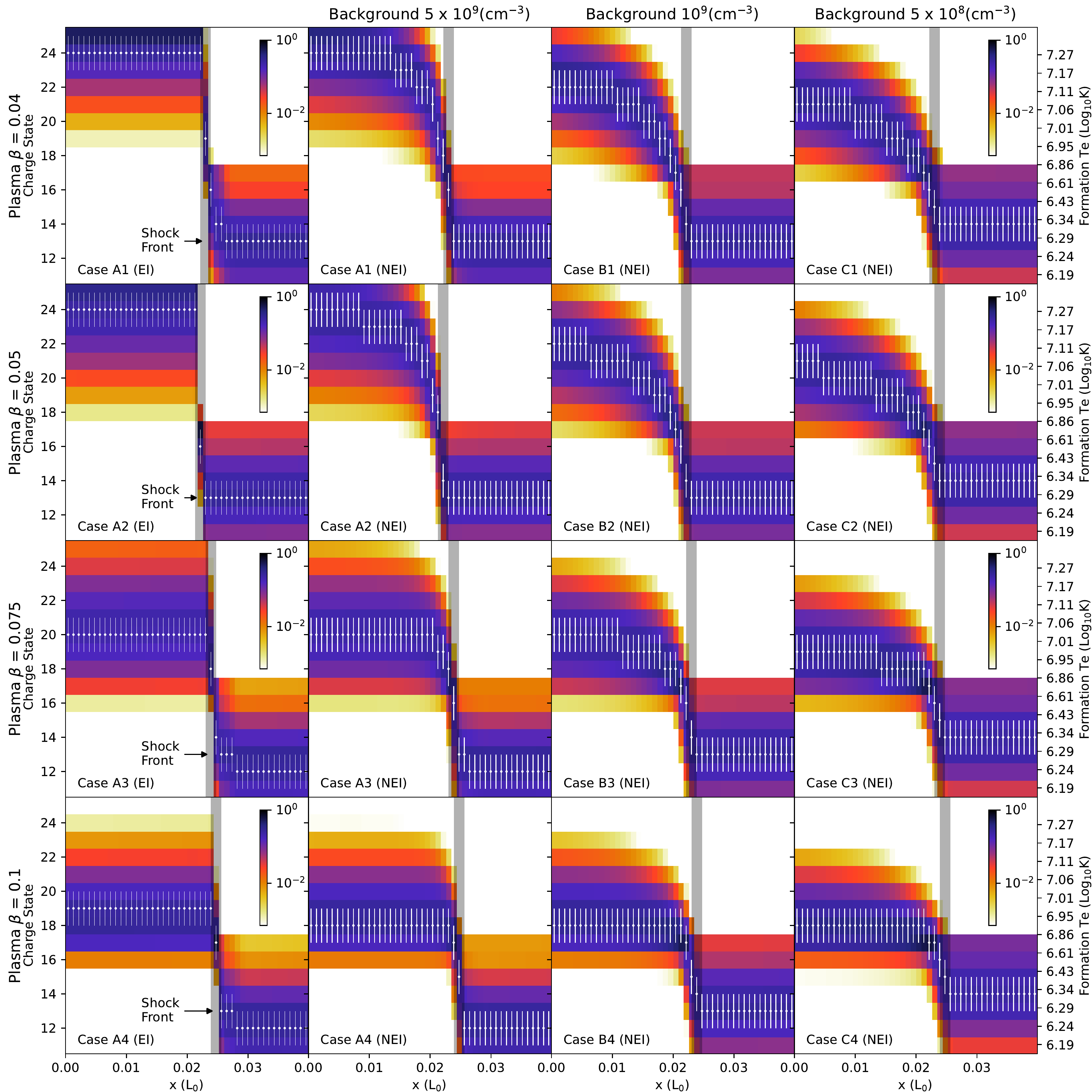}
    \caption{Fe ion fractions along the sampling line (y=-0.9L$_0$, as same as in Figure \ref{fig:13gh_halo_vs_nohalo}(a)) in different cases. 
    The left panel shows the equilibrium ionization results of Case A series, and the right three panels are for NEI situations with different background coronal density ${5 \times 10^9}$, ${10^9}$, and ${5 \times 10^8}$ cm${^{-3}}$, respectively. From top to bottom, each row is for the different plasma $\beta$, 0.04, 0.05, 0.075, and 0.1, which are relative to the post-shocked plasma temperatures around 18 MK, 16 MK, 10 MK, and 8 MK. 
    The white vertical bars indicate the most dominant top three ions along the sampling line at each position. 
    %The right axis labels the corresponding temperature of each ion with the abundance ion population in equilibrium ionization assumptions.
    The right axis labels the corresponding formation temperature of each ion assuming ionization equilibrium.
    }
    \label{fig:3}
\end{figure}

To investigate how plasma temperature and density impact the ion fraction distribution across the shock front, we compare NEI features in detail for different simulation cases.
In Figure \ref{fig:3}, we show the ion population of dominant Fe ions along the horizontal sampling line (dotted lines at $y=-0.9$L$_0$ in Figure \ref{fig:13gh_halo_vs_nohalo}).
The white vertical bars indicate the dominant three charge states at each location along this sampling line, and the background colors are for ion fractions.
In equilibrium ionization, as shown in the left panels in Figure \ref{fig:3}, the Fe ion fractions are dominated by the local temperature distribution, which is roughly uniform along the sampling line inside the reconnection current sheet due to thermal conduction.
Therefore, the EI ion population of each Fe charge state would be the same along the $x-$ direction.
For example, the dominant three Fe ions are shown by dark purple colors on the top row in Figure \ref{fig:3} where the shocked temperature is higher, $\sim 1.8 \times 10^7$ K, inside the current sheet.
Outside the current sheet, the coronal temperature is about 2 MK and the corresponding main ion charge states are \ion{Fe}{12} $\sim$
\ion{Fe}{14} along the sampling line, except near the shock fronts, where the background plasma may be slightly heated by compression in different plasma $\beta$ environments during the magnetic reconnection process.

The right panels of Figure \ref{fig:3} show the NEI ion fraction along the sampling line when the reconnecting current sheet evolves to a steady phase and the density in the downstream side jumps to roughly ${1.1 \times 10^{10}}, 2.3 \times 10^9, 1.3 \times 10^9$ cm${^{-3}}$ for Cases A1, B1, and C1, respectively.
Because the ionization timescale is inversely proportional to the electron density, the tenuous density is expected to cause more NEI effects. 
As shown in the first row of Figure \ref{fig:3}, the dominant Fe ion quickly changes from \ion{Fe}{13} to \ion{Fe}{24} in high density Case A1 in the postshock region. In the low density Case C1, the dominant Fe ion changes more slowly and finally approaches \ion{Fe}{21}, which causes broader distributions of relatively low charge states (e.g., \ion{Fe}{14} $\sim$ \ion{Fe}{20}) in the reconnection current sheet compared with the high-density case. 

Another feature is that the effect of NEI is not only sensitive to the electron density, but also substantially depends on the target temperature: the hot and tenuous plasma shows strong NEI features inside the current sheet compared with relatively dense and cooler plasma.
The higher charge states generally will tend to show significant departures from the EI.
For example, the dominant ion is \ion{Fe}{18} when the temperature is about 8 MK as shown in the fourth row (Cases A4, B4, and C4).
Though the ion fraction changes much more slowly in low-density Case C4 and the higher charge states (e.g., \ion{Fe}{20} $\sim$ \ion{Fe}{25}) show clear differences from EI, the lower charge states (\ion{Fe}{17} $\sim$ \ion{Fe}{19}) are closer to EI except in a narrow region near the shock front,
because ionization rates of Fe ions increase more dramatically as the temperature increases, while the recombination rates change relatively slowly. The imbalance between the ionization and recombination process becomes stronger in higher temperature cases.
%This behavior causes the observed temperature of the reconnection current sheet to be much lower than its real temperature. For instance, the dominant Fe XXII ion indicates a plasma with $\log T = 7.11$ K in NEI case while the `real' plasma temperature can be higher than $\log T= 7.25$ K as shown in Figure \ref{fig:3}. 
The under-ionized nature also causes the temperature of the reconnection current sheet that is usually derived under the EI assumption to be lower than its actual temperature.
For instance, the reconnection current sheet with dominant emission from \ion{Fe}{22} would be thought to be $\log T = 7.11$ K in equilibrium ionization assumptions. However, the `actual' plasma temperature can be higher than $\log T= 7.25$ K, as shown in Figure \ref{fig:3}.

% Figure: difference between NEI fraction and EI fraction
\begin{figure}
    \centering
    \includegraphics[width=0.7\textwidth]{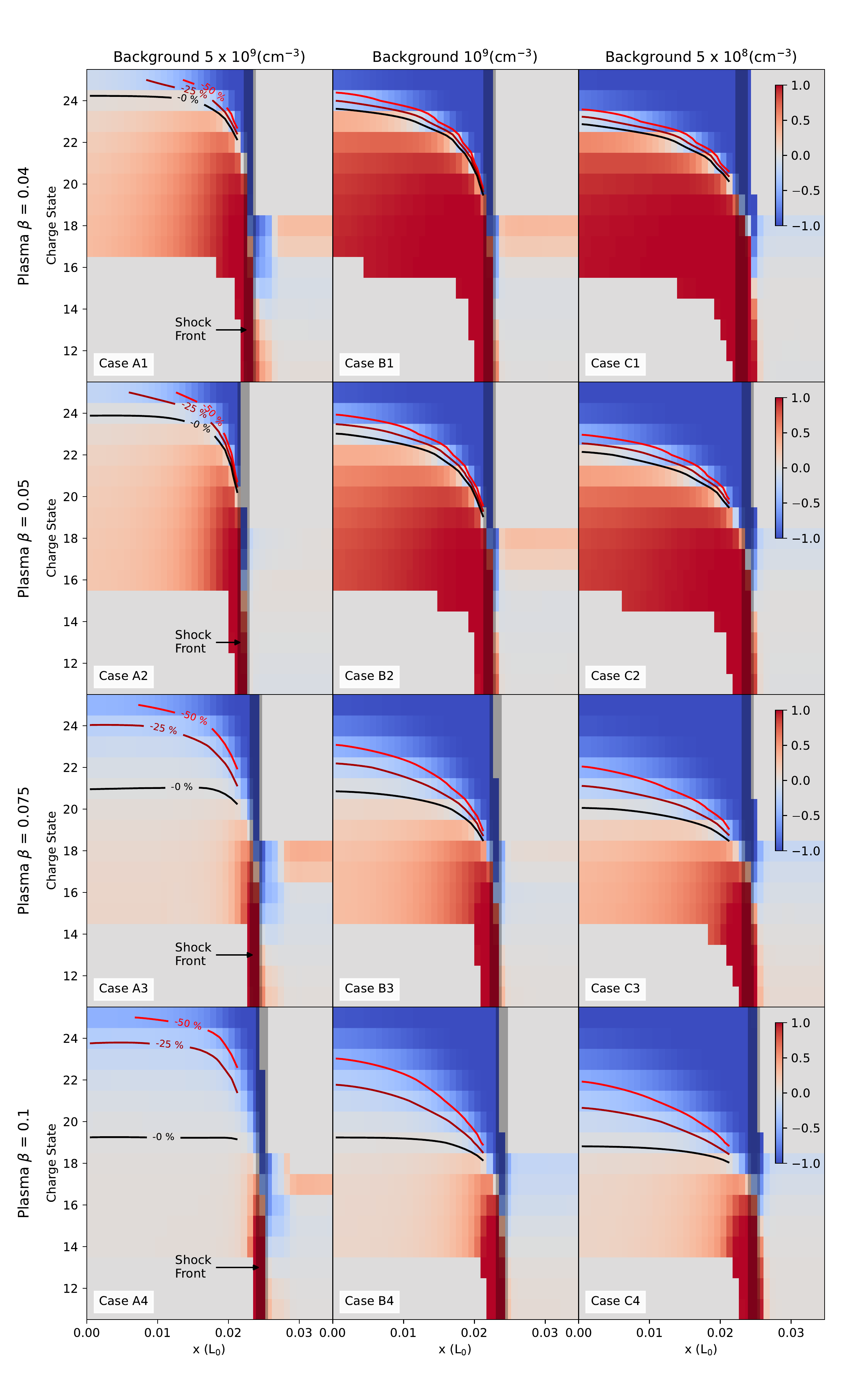}
    \caption{The relative difference of Fe ion population between NEI and equilibrium ionization, defined by ${f_{diff} = (f_{nei} - f_{ei})/(f_{nei} + f_{ei})}$. The colored solid lines marked the position with $f_{diff}$=0, -25\%, and -50\%.
    } 
    \label{fig:4}
\end{figure}

To compare the results of NEI with the ionization equilibrium assumption, we define the relative difference of ion populations between NEI and EI assumption as ${f_{diff} = (f_{nei}-f_{ei})/(f_{nei}+f_{ei})}$, and plot out $f_{diff}$ maps for each case in Figure \ref{fig:4}. 
Here, $f$ is for ion fractions, $nei$ stands for NEI, and $ei$ stands for equilibrium ionization.
$f_{diff}$ is a compressed scale to show departures in two directions, which gives +1 when the equilibrium ion fraction is close to zero and the non-equilibrium ion fraction is close to 1, and vice versa for ${f_{diff}=-1}$. 
The relative value of populations then can be obtained by,
\begin{equation}
    \frac{f_{nei}}{f_{ei}} = \frac{(1+f_{diff})}{(1-f_{diff})}.
\end{equation}
For instance, $f_{diff} = 0.25$ means that $f_{nei} \sim 1.67f_{ei}$, and $f_{diff} = 0.5$ corresponds $f_{nei} \sim 3f_{ei}$, respectively.

The format of Figure \ref{fig:4} is similar to Figure \ref{fig:3}, and the color maps range from -1 to 1 accordingly. 
The black contour lines denote the position where the NEI population is equal to the EI assumption with $f_{diff} = 0$, and the contours of $f_{diff} = -25\%, -50\%$ are marked by dark red and red lines, respectively.
Due to the under-ionized nature inside the current sheet, the high charge state ions (above the red lines in Figure \ref{fig:4}) lie in the blue region that indicates that the fractions in NEI are much smaller than in EI, while the low charge state ions are red because the factions in NEI are larger than in EI.
For dense plasma situations (e.g., Cases A1--A4 in Figure \ref{fig:4}), the fraction of high charge states obviously departs from EI, especially in the local downstream region near the shock front. Even in the high-temperature and high-density environments (Case A1), the relative difference ($f_{diff}$) of \ion{Fe}{25} can be as high as 50\% in a relatively wide region ($\sim$ $1.2 \times 10^6$m from ${x=0.015L_0}$ to ${x=0.023L_0}$, with $L_0 = 1.5 \times 10^8$m).
If the background density is lower, $\sim 5\times10^8$ cm$^{-3}$ (Cases C1-C4), all high charge state ions clearly depart from EI assumptions.

On the other hand, the relative differences $f_{diff}$ are much smaller around the equilibrium ionization position (black lines) in low-temperature ranges.
For instance, in Case A4, the equilibrium ionization ions are dominated by \ion{Fe}{19} while ions ranging from \ion{Fe}{14} to \ion{Fe}{23} are close to EI states, with the $f_{diff} < \sim$ 25\%. 
This suggests that the shocked plasma can be close to equilibrium ionization states for lower temperature current sheets (e.g., 8 MK) even if the density is lower to $\sim 5 \times 10^8$ cm$^{-3}$. 
In this case, the ionic fraction of lower ionization ions, such as \ion{Fe}{14} and  \ion{Fe}{18}, can be estimated based on EI results.
However, the reconnection current sheet of higher temperature (e.g., 18 MK and above) with low densities should be considered as in NEI states.  Inside such a current sheet,  the population of most observable ions (e.g., \ion{Fe}{18} to \ion{Fe}{24}) may significantly depart from EI assumptions up to about a factor of three.

\subsection{Shock Fronts in Synthetic EUV images}
The high-temperature plasma sheets observed by EUV imaging instruments (e.g., SDO/AIA) above the post-flare loops are commonly thought as the appearance of reconnecting current sheets. Some bright current sheet features observed in the SDO/AIA 94 and AIA 131 bands show a very similar morphology compared with theoretically predicted reconnecting current sheet structures. 
However, systematic analysis of how NEI affects these EUV imaging observations is still rare. In particular, the EUV emission around Petschek-type shock fronts with NEI effects is not fully understood in previous theoretical modeling studies.        
Therefore, we calculate emission intensities using the NEI modeling results and obtain synthetic SDO/AIA images around Petschek-type shocks. 
Here, we include 14 elements that contribute significantly to coronal emission in the combined MHD-NEI simulations, including H, He, C, N, O, Ne, Mg, Al, Si, S, Ar, Ca, Fe, and Ni. For each element, the emission intensities are calculated using the NEI ionic population and the emissivity data from the atomic database, CHIANTI \citep{Dere_2019ApJS..241...22D}. 
For each ion of a chosen element, the line emission can be calculated by:
\begin{equation}
    I_{line}(\lambda) = \frac{n_x}{4\pi n_H}\int {G(T)} \,{\rm d}EM(T),
\end{equation}
where $I_{line}$ is in units of photon cm$^{-2}$ s$^{-1}$ sr$^{-1}$, $\lambda$ is the wavelength, and $n_x/n_H$ is the elemental abundance \citep{Schmelz2012ApJ...755...33S}. $G(T)$ is the contribution function that is calculated by using ion fractions in NEI states and emissivities are calculated by using the $emiss\_calc$ package in CHIANTI database. $EM$ is the emission measure, assuming the line-of-sight (LOS) of $10^9$ cm.
The SDO/AIA effective areas are applied to get count rates in each AIA band. Finally, the AIA intensity that counts all emission lines from the above 14 elements is calculated at each cell of the MHD simulations.

\begin{figure}
    \centering
    \plotone{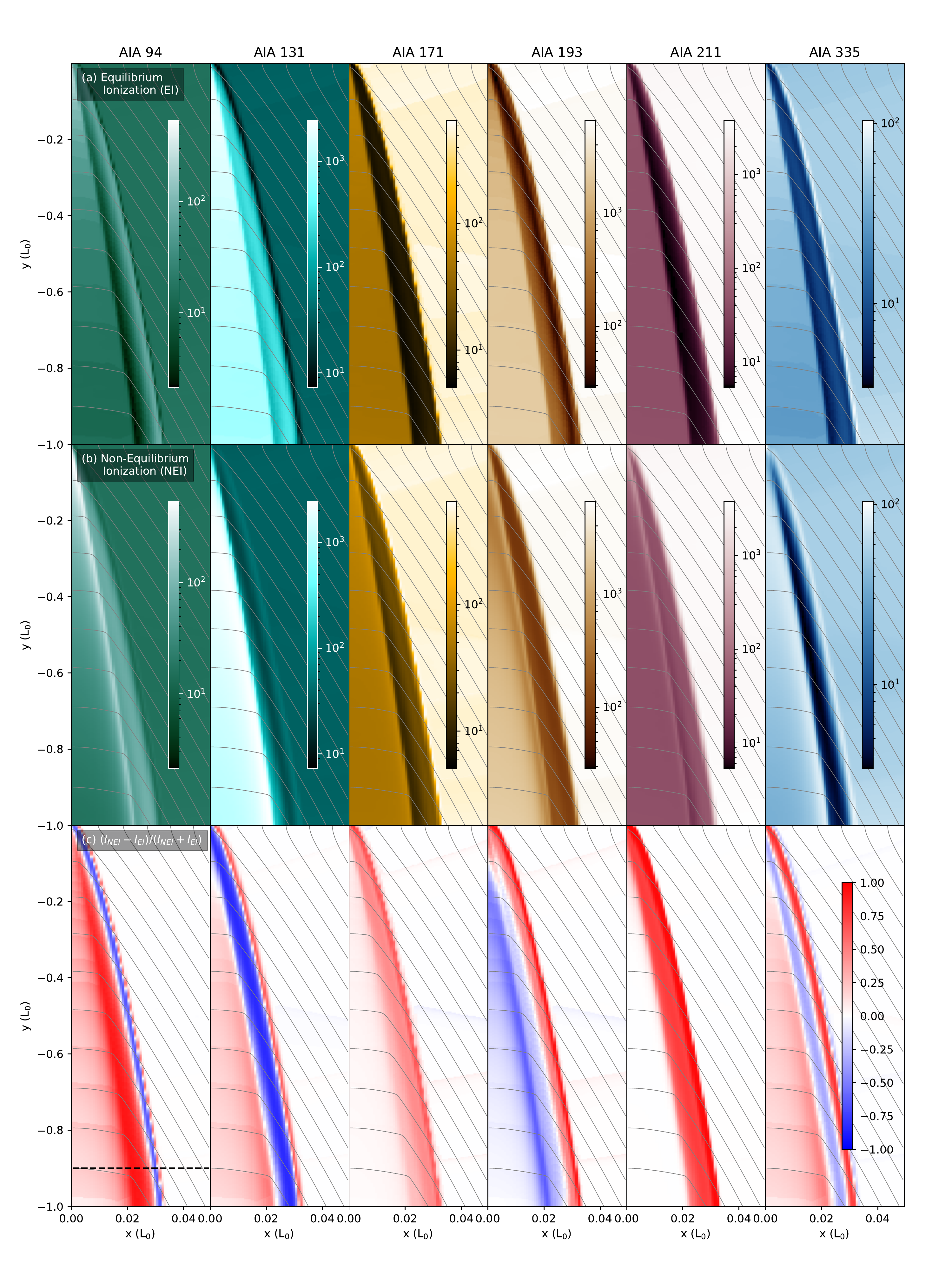}
    \caption{Synthetic SDO/AIA images for Case A2$^*$ with classical Spitzer thermal conduction coefficient.
    Row (a) and (b) are total count rates (DN s$^{-1}$ pixel$^{-1}$) for the equilibrium ionization and NEI results, respectively.
    (c) shows the relative difference ${(I_{nei} - I_{ei})/(I_{nei} + I_{ei})}$ between NEI and EI count rates.}
    \label{fig:aia_halo}
\end{figure}

Figure \ref{fig:aia_halo} shows the synthetic SDO/AIA count rate for Case A2$^*$ with the thermal halo structures outside the shock front. The first row is AIA intensities for the equilibrium ionization assumption and the second row is results using the non-equilibrium ionization results, respectively.
% 1. bright center region + NEI
Because the postshock temperature is around $1.6 \times 10^7$ K (see Figure \ref{fig:13gh_halo_vs_nohalo}(b)), an overall feature is that the inside region of the reconnection current sheet appears bright on high-temperature AIA channels, such as AIA 94 and AIA 131, in both NEI and EI cases. Other AIA channels show lower count rates in the current sheet than ambient plasma.
%% 2. The halo regions
Under the EI assumption, the thermal halo can be clearly recognized in all SDO/AIA channels because the emission intensity is only temperature dependent (Figure \ref{fig:aia_halo}(a)). It appears as significantly low emission sheathes outside the Petschek-type shock fronts in AIA 94, 171, 193, 211, and 335 channels. 
For the higher temperature AIA 131 channel, the thermal halo shows an extending bright edge around the shock fronts. 
However, the NEI synthetic images show that the signal of thermal halo regions is significantly weaker than with the EI assumption. As shown in Figure \ref{fig:aia_halo}(b), the low emission sheath is hard to observe in AIA 131, which causes a narrower bright reconnection region compared to the EI image in this channel.
In contrast, the AIA 94 and AIA 211 maps generally show a wider bright reconnection current sheet region compared with EI images due to the contribution of NEI effects on thermal halo regions.
%% 3. The relative differences
We can see the NEI effects on intensities from the scaled relative difference (${(I_{nei} - I_{ei})/(I_{nei} + I_{ei})}$) of the synthetic emission intensity between NEI and EI cases (see the third row of Figure \ref{fig:aia_halo}). 
Inside the reconnection region, the EI assumption causes underestimated count rates in the AIA 94, 131, and 335 channels, but gives a higher AIA 193 intensity compared with NEI results. Around the shock front and the thermal halo regions, AIA 131 count rates under the EI assumption are clearly over-estimated compared to NEI predictions, while AIA 171 and 211 are under-estimated.
%[I am confused by the use of underestimated and overestimated.  I would say that EI underestimates things if NEI is higher, since NEI is 'correct'.  If I understand these sentence, 'AIA 131 intensity can be underestimated' really means that NEI predicts a lower brightness than EI.]

\begin{figure}[htb!]
    \centering
    \plotone{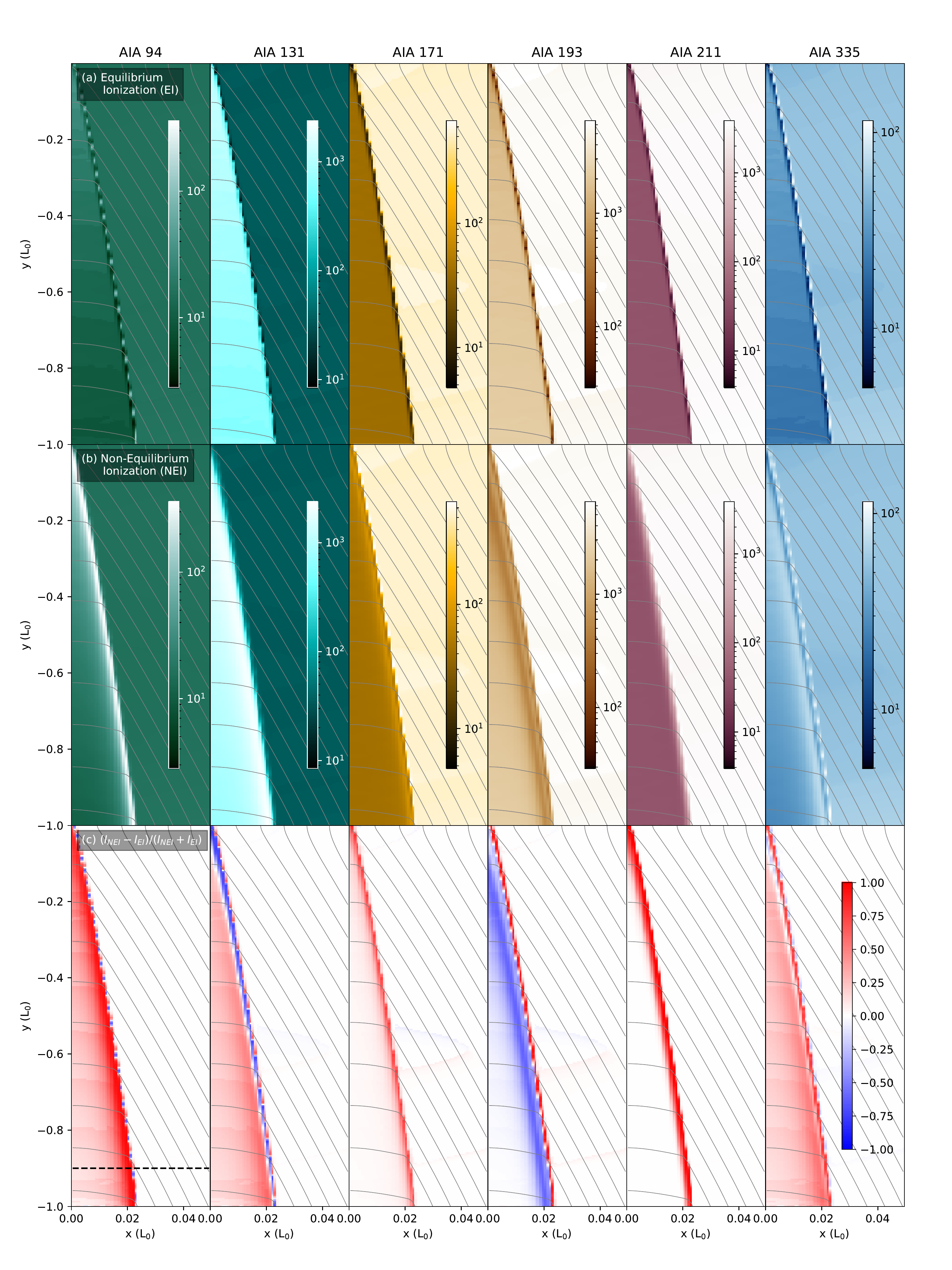}
    \caption{Synthetic SDO/AIA images for Case A2 applying conduction flux-limitation. The panels are defined as same as in Figure \ref{fig:aia_halo}. }
    \label{fig:aia_fluxlimt}
\end{figure}

%% Conduction Flux limited cases
As a comparison, we show the synthetic SDO/AIA images in Figure \ref{fig:aia_fluxlimt} for Case A2 with thermal conductive flux-limitation. The  under-estimated AIA 94, 131, 171, and 335 features in equilibrium ionization assumptions and over-estimated AIA 193 intensity are all as the same as in Case A2$^*$.
Because there is no clear thermal halo in this model, the EUV intensity clearly jumps across the shock front as shown on most bands such as AIA 171, 131, and 193 due to both the temperature and density jumps.
%[Mention density change of gas at the shock?]
Furthermore, the NEI also causes an enhanced brighter edge close to the shock front on the post-shock side (Figure \ref{fig:aia_fluxlimt}(b)(c)) which is generally brighter than the thermal halo regions.
Unlike the thermal halo situation, the reconnection current sheet can be slightly narrower in AIA 211 due to the NEI effects.

\begin{figure}[htb!]
    \centering
    \plotone{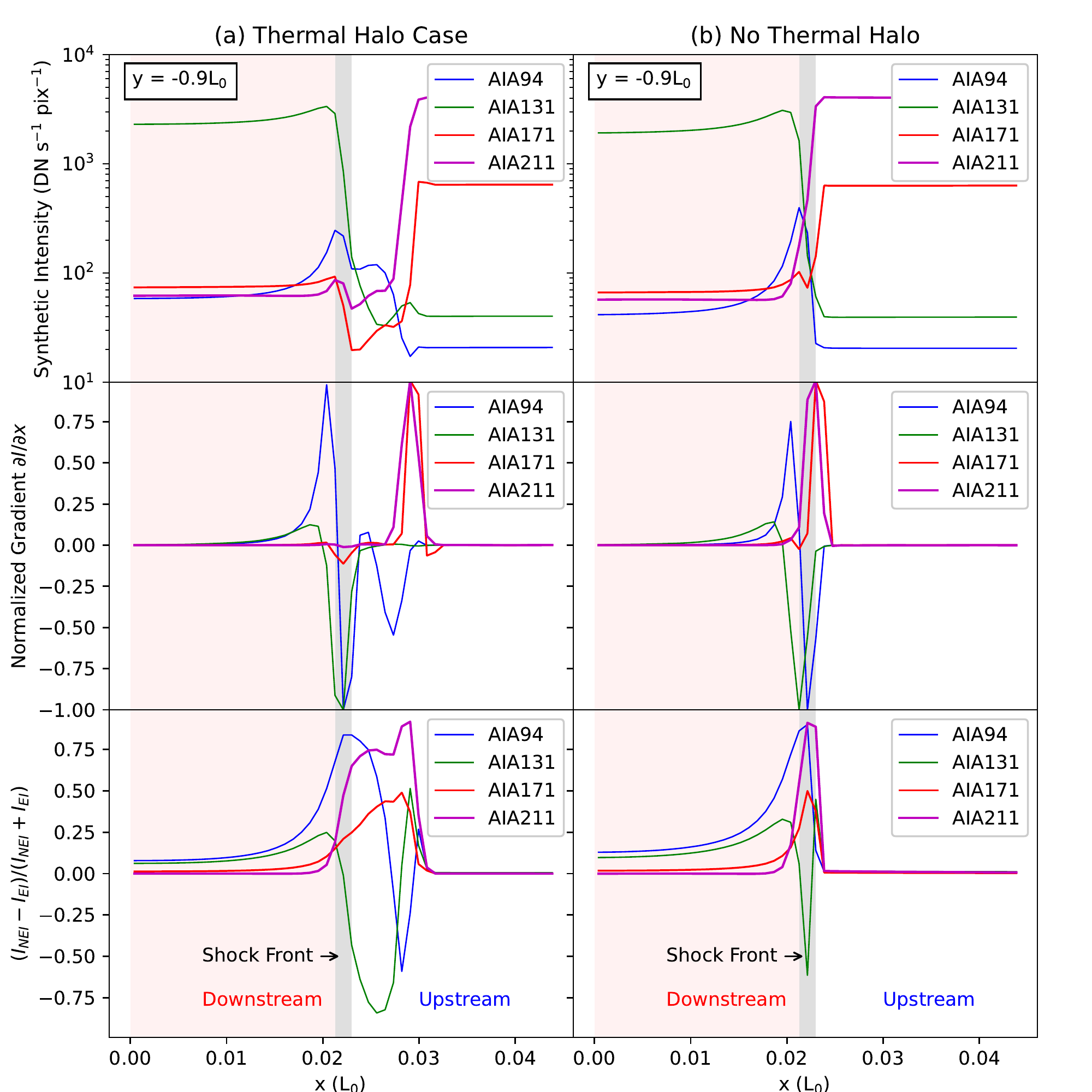}
    \caption{Synthetic SDO/AIA intensity in NEI around the shock front along the horizontal sampling line at $y=-0.9L_0$ as shown by dashed line in Figures \ref{fig:aia_halo} and \ref{fig:aia_fluxlimt}. The left panels (a) are for Case A2$^*$ with the thermal halo, and the right panels (b) are for Case A2 without the thermal halo, respectively. The top rows are emission intensity ($I$), the middle rows are normalized intensity gradient in $x-$ direction ${\partial I/\partial x}$, and the third row are predicted intensity difference between NEI and EI results, ${(I_{nei} - I_{ei})/(I_{nei} + I_{ei})}$.}
    \label{fig:aia_line_compar}
\end{figure}

%% Short summary: a94, 131, 171, 211
Here we look into the details of emission from the thermal halo and shock front along a chosen sampling line at the height $y=-0.9L_0$ (as shown by dashed lines in Figures \ref{fig:aia_halo}, \ref{fig:aia_fluxlimt}). Figure \ref{fig:aia_line_compar} shows the predicted intensity profiles for AIA 94, 131, 171 and 211 bands.
In the model with a thermal halo (Figure \ref{fig:aia_line_compar}a), the intensity jump for high-temperature channels, such as AIA 131, mainly appears at the shock front (indicated by vertical gray shadows). On the other hand, the relatively low temperature AIA channels (e.g., 171 and 211) show clear jumps at the edge of the thermal halo region.
The above intensity changes are more easily seen by plotting the normalized gradient profiles in the second row of Figure \ref{fig:aia_line_compar}. The dominant AIA 171 and 211 gradient peaks appear at the thermal front ($x \sim 0.03L_0$), while the negative gradient peak of AIA 94 and 131 appears at $x \sim 0.023L_0$.
In comparison, in the model without the thermal halo, all dominant AIA intensity jumps can be found at the shock front where the gradient peaks of AIA 94 and 131 are still negative, and AIA 171 and 211 are positive, respectively.

To further improve our understanding of how NEI affects the SDO/AIA intensity changes crossing the shock front, we plot out the relative difference between NEI and EI in the third row of Figure \ref{fig:aia_line_compar}.
In the thermal halo case, the EI assumption under-estimated the AIA 171 and 211 intensities in both the downstream and thermal halo regions, and over-estimated the AIA 131 in the thermal halo region by more than a factor of three.
%[It is better to say that EI under- or over-estimates the intensities, since  NEI is 'correct'.]
In conclusion, the analysis of different variations of EUV intensity around the shock front serves as a potential diagnostic tool for understanding the shock properties. From the pre-shock plasma to the downstream of slow mode shocks, one can expect the positive AIA 171/211 jump following with a minimum value on the AIA 94/131 gradient. The gap between the two jumping peaks may indicate the halo region.

It is interesting to analyze how NEI affects the plasma temperature derived from images in the above six SDO/AIA channels.
In general, the emission measure (EM) reconstruction method is used to investigate the plasma temperature and density in high-temperature plasma during solar eruptions. The differential emission measure (DEMs) can be calculated from the intensities of six SDO/AIA channels (94, 131, 171, 193, 211, and 335 \AA) that are dominated by emissions from iron lines and other elements \citep{O'Dwyer_etal_2010A&A...521A..21O}, and the DEM weighted average temperature can be used to estimate the temperature of reconnection current sheets. 
However, NEI may cause a significant departure in the DEM weighted temperature from the `real' plasma temperature. We, therefore, perform DEM analysis based on the above synthetic AIA images and make detailed comparisons in different cases.

In observational studies, the DEM-reconstructed temperature strongly depends on the multiple temperature model which may cause slightly different results with different solvers \citep[e.g.,][]{Cheung2015ApJ...807..143C,Szenicer2019SciA....5.6548S}.
In our case, the modeled synthetic SDO/AIA intensity can be thought as a $single-T$ component profile: it is integrated along the $z-$ direction where the temperature is assumed to be uniform along the LOS. The DEM distribution can be obtained by using the $single-T$ approximation.
We then simply estimate the reconstructed temperature $T_{nei}$ according to the $\chi^2$ minimization method by setting
\begin{equation}
    \chi^2(x,y,T) = \sum_{i=1}^6 (I_{nei}(x,y)_i/n_e(x, y)^2 - I_{ei}(T)_i)^2     .
\end{equation}
Here $I_{ei}(T)_i$ is the intensity of the $i$th channel among six AIA bands with the unit plasma density in equilibrium ionization assumptions, and $n_e$ is the density in MHD models.

\begin{figure}[htb!]
    %% Case 7 and 8
    \centering
    \includegraphics[width=0.65\textwidth]{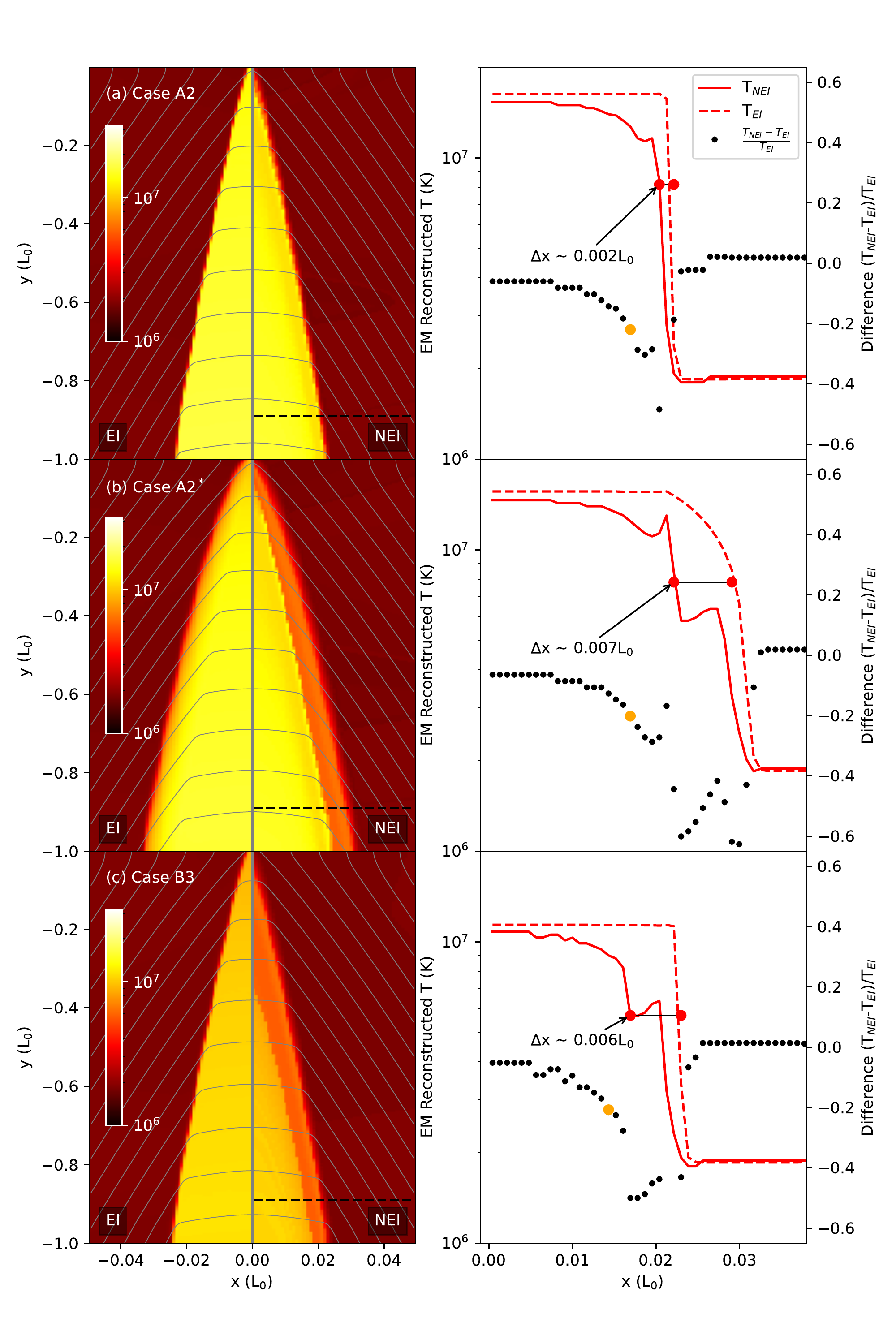}
    \caption{Emission measure reconstructed temperature distribution using the synthetic SDO/AIA intensity for Case A2, A2$^*$, and B3. 
    In each temperature map, the left region ($x$ ranges from -0.05 to 0) shows the MHD modeled temperature (or temperature in EI), and the right part ($x$ ranges from 0 to 0.05) is for the reconstructed temperature. 
    The right panels are temperature distribution, and the relative difference between NEI and EI results along the horizontal sampling line at $y=-0.9$L$_0$ (dashed black lines on the left panels). 
    The two red dots mark the location of the half-maximum temperature, and $\Delta$x indicates the distance between the above two red cycles in $x-$ direction. The orange dot indicates the position with $\sim$ 20\% difference along the dotted lines.}
    \label{fig:aia_emTe}
\end{figure}

Figure \ref{fig:aia_emTe} compares the reconstructed temperature distribution with the MHD modeled temperature in three cases. Figure \ref{fig:aia_emTe}(a-b) are for the Cases A2 and A2$^*$ where the shocked plasma is heated to $\sim 1.6\times 10^7$ K with density higher than $\sim 10^{10}$ cm$^{-3}$, and Figure \ref{fig:aia_emTe}(c) is for the relatively lower temperature ($\sim 10^7$ K) and density ($\sim 2.5 \times 10^9$ cm$^{-3}$) situation in Case B3.
Consistent with the above NEI analysis, the under-ionized feature inside the reconnection current sheet causes a significant underestimation of the EM reconstructed temperature. 
As shown in the temperature maps in Figure \ref{fig:aia_emTe}, the high-temperature plasma sheet appears narrower than the actual high-temperature plasma sheet in all three cases. 
Solid red lines clearly show the lower reconstructed temperature along the horizontal sampling lines. We also plotted out the relative difference (${(T_{NEI}-T_{EI})/T_{EI}}$) between reconstructed temperature ($T_{NEI}$) and real one ($T_{EI}$) on the right panels in Figure \ref{fig:aia_emTe}. The relative difference can be as high as 60\% which occurred near the shock front. The orange dots marked the location where the reconstructed temperature is noticeably lower than the actual temperature, $\sim$ 20\% at $x\sim0.017$ in Case A2 (and A2$^*$) and $x\sim0.015$ in Case B3. 
In Figure \ref{fig:aia_emTe}, The red dots mark the edge of a high-temperature current sheet by using the half-maximum temperature, and the distance between two red dots from NEI and EI profiles is shown by $\Delta x$ accordingly. We can see that the apparent width of current sheets may be narrower than their actual width by about $8\%, 31\%$, and $\sim 36\%$ according to the emission reconstruction method in the above Case A2, A2$^*$, and B3, respectively.

An expanding halo region can be found in Case A2$^*$ shown by Figure \ref{fig:aia_emTe}(b). The reconstructed low-temperature edge comprises two parts: the thermal halo region due to thermal conduction and the lower $T_{NEI}$ regions due to NEI. Compared with Case B3 (without the thermal halo), it is interesting that both reconstructed low-temperature edges show very similar distribution features. Therefore, the diagnostic of shock structures and possible thermal halos on high-temperature emission requires the analysis of NEI effects in detail.

%========================================
\subsection{Reconnection Current Sheet above Closed Loops}
%========================================
\begin{figure}[htb!]
    \centering
    \plotone{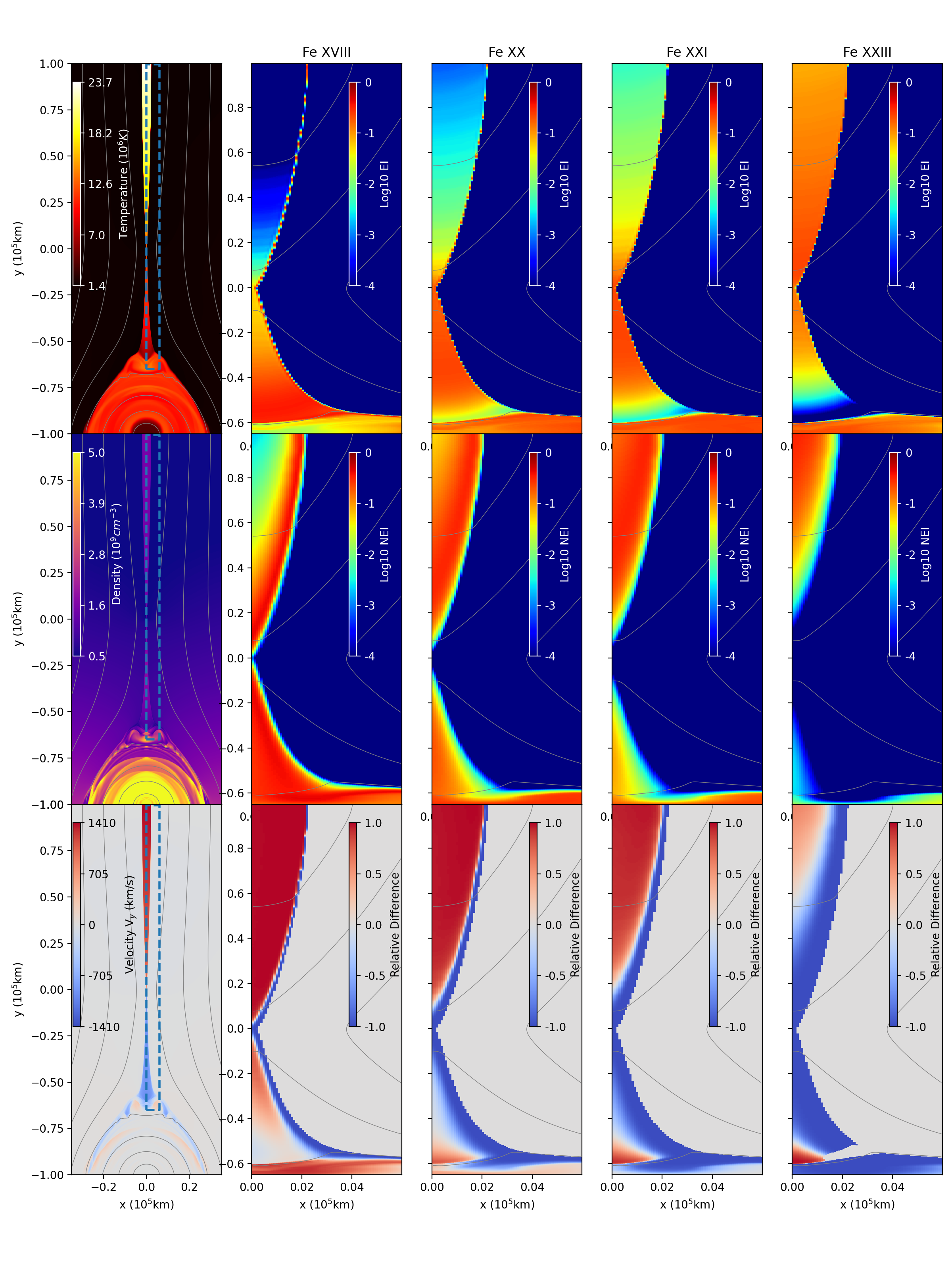}
    \caption{Ionization distribution with the line-tied boundary condition in Case D. The left panel shows the distributions of temperature, density, and velocity component $V_y$ at the time $t=16t_0$.
    The right panels are ion fraction of \ion{Fe}{18}, \ion{Fe}{20}, \ion{Fe}{21}, and \ion{Fe}{23} in the local region shown by the blue box on the left panels. The top row is ion fractions in the equilibrium ionization ($f_{ei}$), the second row is for NEI ($f_{nei}$), and the third row shows relative differences between NEI and EI defined by $(f_{nei}-f_{ei})/(f_{nei}+f_{ei})$.
    }
    \label{fig:linetied_var}
\end{figure}

In this section, we perform a more realistic simulation following the configuration of the classical solar flare model \citep[e.g.,][]{Kopp-Pneuman1976}. In this case (see detailed simulation parameter in Table \ref{tab:parameters}, Case D), we apply a similar setup except introducing gravity in the $y-$ direction, and utilizing the line-tied boundary condition at the bottom to ensure that the magnetic field lines are rooted on the solar surface. 
Driven by the initial perturbation on magnetic fields \citep[e.g.,][]{Shen_2018ApJ...869..116S}, the magnetic reconnection quickly takes place in the initial current sheet, and a pair of reconnection outflows appear along the vertical direction ($y-$) as the same as in Case A2. However, the reconnected magnetic flux gradually accumulates at the bottom due to the line-tied condition, and the closed magnetic loops appear at the bottom representing the flare loops. 
In this case, the downwards moving outflow collides with the newly closed magnetic loops and may form termination shocks if the outflow speed exceeds the local fast-magnetosonic speed in the loop-top region \citep{Forbes1986ApJ...305..553F}.
In general, these closed magnetic loops also cause asymmetrical reconnection outflow behaviors along the reconnection current sheet: the downwards outflow generally has less speed compared with the upwards one.
Therefore, the ionization states could be significantly different along the reconnection current sheet.

Figure \ref{fig:linetied_var} shows the spatial distribution of the primary plasma variables (temperature, density, $V_y$) and the ion fractions of selected Fe ions in the reconnection current sheet and flare loops regions. At this time ($t=16 t_0$), the reconnection evolved into the relatively steady phase when the flare loops already fully developed. As the same as in other cases in Table 1, the enhanced resistivity diffusion center is still located at the system center ([$x=0, y=0$]) at which the primary reconnection X-point can be found.
Bi-direction high-temperature reconnection outflows are displayed in the left panels of Figure \ref{fig:linetied_var}. Here, the upward outflows can reach $2 \times 10^7$ K with the maximum flow speed exceeding 1400 km/s. On the other hand, the temperature of the downwards reconnection jet is as low as $\sim 10^7$ K and its velocity is relatively slow ($\sim$ 800 km/s). 
In equilibrium ionization assumptions, high ionization ions (e.g., \ion{Fe}{23}) should mainly dominate the upward outflow due to the extremely high temperature, and more \ion{Fe}{18} and \ion{Fe}{20} ions could appear in the downward outflows because of the slightly lower temperature as shown in the top row of Figure \ref{fig:linetied_var}.
However, we will discuss how is the actual ion distribution different from the EI assumption.

The NEI effect can cause substantial departures of the ion distribution from equilibrium ionization states in both upwards and downwards outflow regions, shown by the second row of Figure \ref{fig:linetied_var}.
Consistent with the above discussion of Petschek-type reconnection configurations (e.g., Figure \ref{fig:3}), the upwards flow displays clear under-ionized features.
The populations of \ion{Fe}{18}, \ion{Fe}{20}, and \ion{Fe}{21} in NEI are all substantial in the upward outflow regions compared with EI assumptions in which, for instance, the \ion{Fe}{18} is rare.
The ion fraction difference between NEI and EI can be large, as shown in the third row of \ref{fig:linetied_var}. For example, the maximum relative difference of \ion{Fe}{18} is close to one along the upwards reconnection flow. The \ion{Fe}{23} also shows the same behavior, except a slightly negative difference appears in the edge region inside the current sheet.

Along the downwards reconnection jet, the ionization state is more complex than the ideal Petschek-type reconnection current sheet due to the dramatic variation of both temperature and density with height. 
As shown in Figure \ref{fig:linetied_var}, the plasma temperature is high near the reconnection X-point, but quickly decreases near the lower tip of the reconnection current sheet because of the expansion of the outflows in the horizontal direction.
The relatively low-temperature plasma continually outflows downwards until it collides with the closed magnetic loops, where the outflow speed drops substantially. In the situation including super-magnetosonic reconnecting outflows, the plasma can be quickly heated again behind the termination shock.
The above temperature variation then causes two different types of NEI states: (i) Near the X-point, the plasma is under-ionized, the same as in the classical Petschek-type reconnection current sheet (e.g., the upwards reconnection jets in this model); (ii) The plasma could be in over-ionized states due to the rapid temperature drop at the lower tip of downward flows.
As shown in the third row of Figure \ref{fig:linetied_var}, the relative difference of \ion{Fe}{18} appears clearly negative values (indicated by blue colors) at the lower end of the reconnection downflows while the other three higher ionized ions become more abundant.  

\begin{figure}[htb!]
    \centering
    \plotone{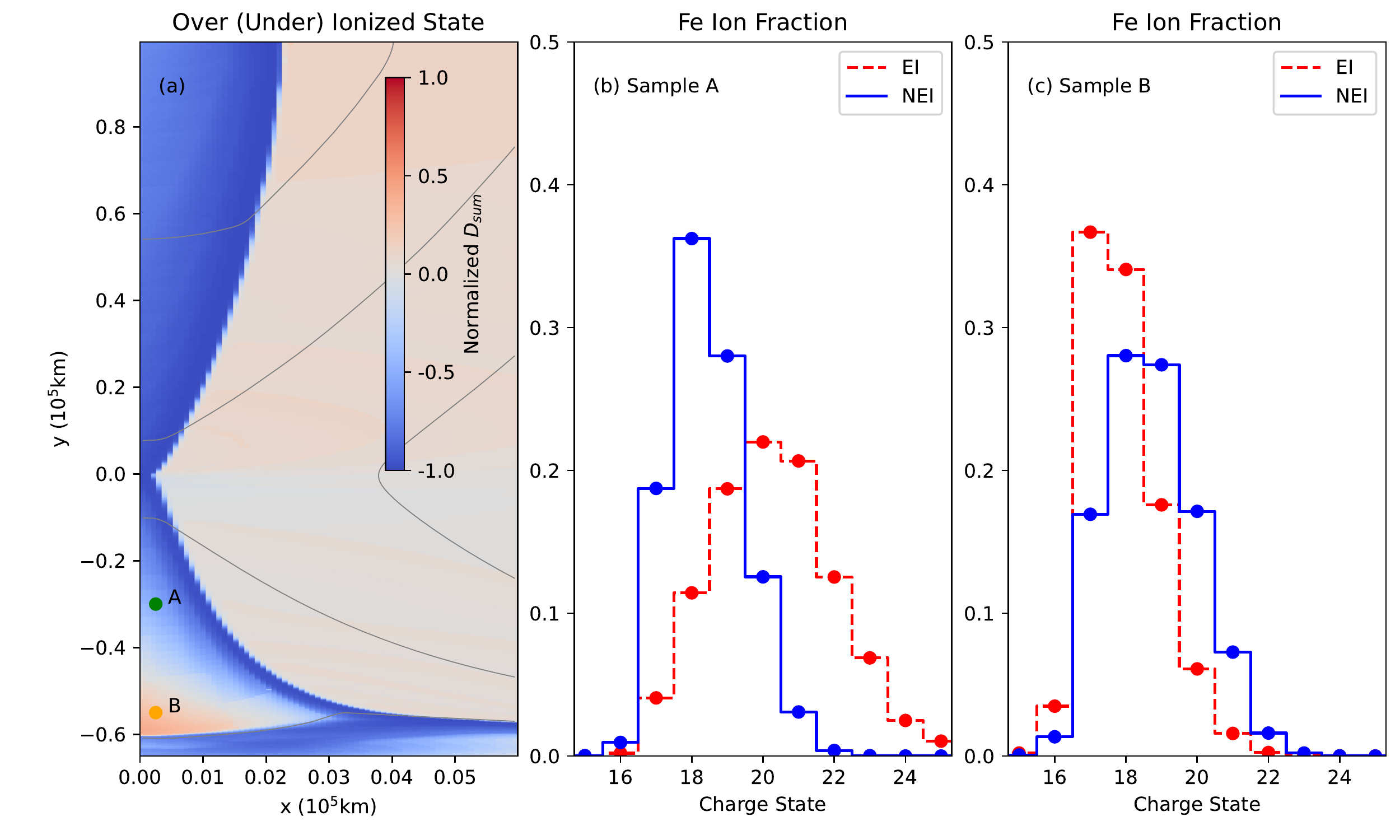}
    \caption{Under-ionized and over-ionized ionization states in downwards outflow regions. Panel (a) shows the normalized total difference ($D_{sum}$) between NEI fraction and EI assumption; (b) and (c) are Fe ion populations at two chosen sampling points indicated by green and orange cycles in panel (a), respectively.}
    \label{fig:oct12d_ionstate}
\end{figure}

The reversal between under-ionized and over-ionized states can be clearly seen in Figure \ref{fig:oct12d_ionstate}. We introduce the normalized total difference of ion fraction to show the departure of NEI from the EI, defined by 
\begin{equation}
    D_{sum} = \sum_{1}^{i_{equal}} -\frac{(f_{nei} - f_{ei})}{2} + 
             \sum_{i_{equal}+1}^{Z+1} \frac{(f_{nei} - f_{ei})}{2}.
\end{equation}
Here $f_{nei}$ and $f_{ei}$ are ion fraction of the $i$th charge state in NEI and EI cases, $Z$ is the atomic number of the chosen element to be calculated, and
$i_{equal}$ is used to point to a particular charge state where the non-zero NEI fraction is equal (or closest) to the EI fraction.
%[How is that possible? Populations add up to one.] 
In this way, $D_{sum}$ can range from  -1 to 1 depending on the ionization state. Under-ionization causes a negative $D_{sum}$ value,  and $D_{sum} > 0$ is for the opposite case when the dominant NEI fractions are larger than EI in over-ionized states.
Figure \ref{fig:oct12d_ionstate} and animation (a) shows the $D_{sum}$ distribution of Fe ions around the reconnection current sheet. The upward outflow is dominated by negative $D_{sum}$ due to the under-ionized plasma, while the under-ionized state could change to the over-ionized state at the low tip of the reconnection current sheet.
Two sampling points (A and B in Figure \ref{fig:oct12d_ionstate}(a)) are chosen to show the Fe ion fraction profile at different heights. As shown in panels (b-c) in this figure, the ion charge state distribution in NEI (solid line) is skewed toward lower charge states compared to equilibrium ionization results (dashed line) at the upper sampling point A, and shifted to higher charge states at the lower sample B.

\begin{figure}[htb!]
    \centering
    \plotone{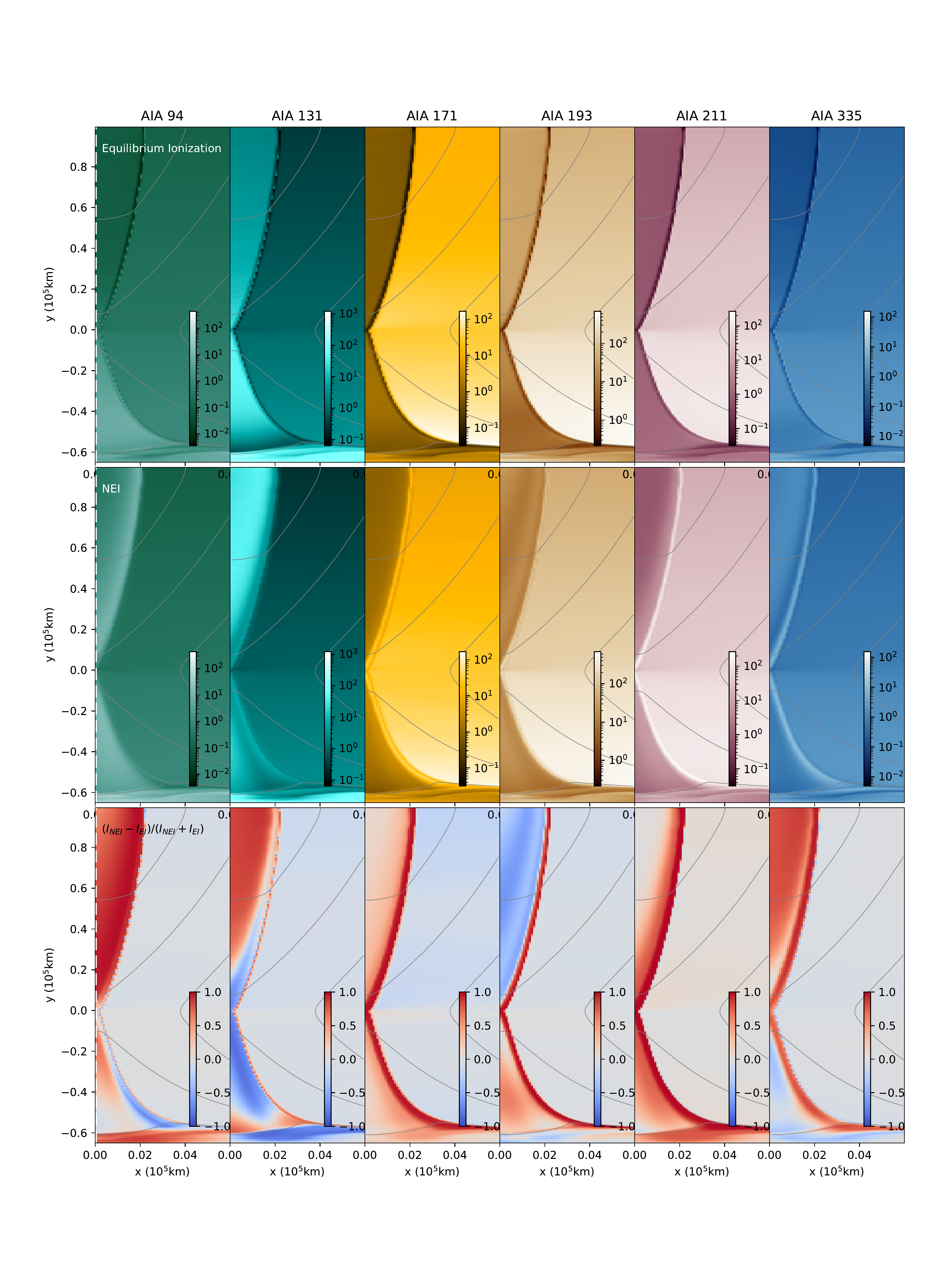}
    \caption{Synthetic SDO/AIA images in Case D applying the line-tied boundary at the bottom. The first row shows AIA count rates in equilibrium ionization assumptions, and the second row is for NEI results. The third row shows the relative difference between NEI and EI results.}
    \label{fig:linetied_aia}
\end{figure}

Figure \ref{fig:linetied_aia} shows predicted SDO/AIA count rate images from the above modeling with the line-tied boundary. The first row is for equilibrium ionization results, and the second row shows the intensities of NEI calculations, respectively.
Similar to the classical Petschek-type reconnection current sheet, the NEI causes stronger emissions in upward outflows regions in AIA 94, 131, 171, 211, and 335 channels. The largest relative difference between NEI and EI count rates appears in AIA 94 and 131 channels, which are close to unity inside the current sheet.
In downward outflows, either higher or weaker NEI count rates can be found at different heights. On the AIA 94 map, the reversed ionization charge states at the lower tip of the current sheet also cause weaker emission (as shown by blue colors in the third row of Figure \ref{fig:linetied_aia}).

\begin{figure}
    \centering
    \plotone{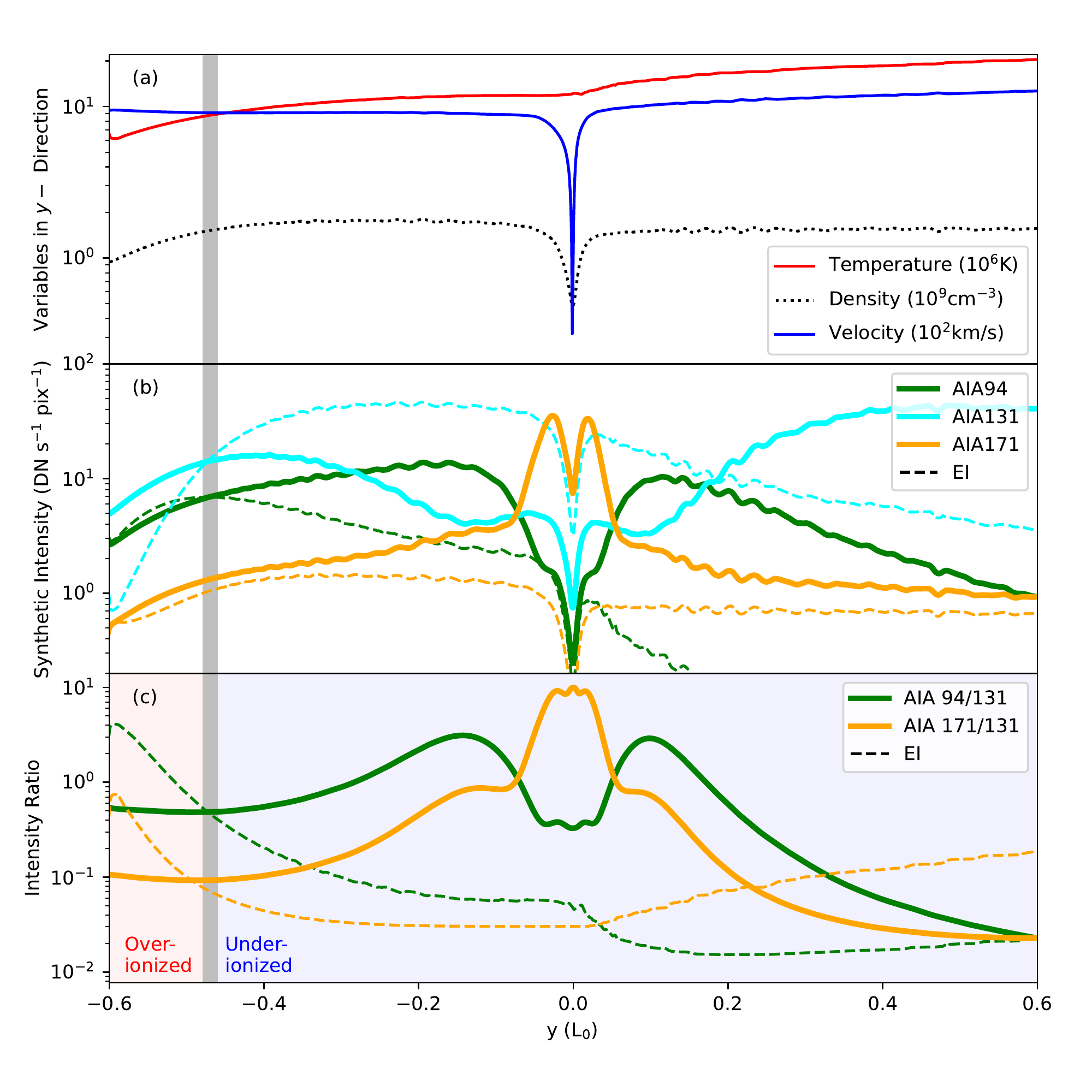}
    \caption{Synthetic SDO/AIA intensity and primary variable distributions along the reconnecting current sheet in Case D.
    (a) Temperature, density, and velocity profiles along the white vertical sampling line in Figure \ref{fig:linetied_aia}. The vertical gray shaded region indicates the position where the ionization property of the reconnection current sheet changes from under-ionized to over-ionized states.
    (b) The chosen SDO/AIA count rates profiles. The bold lines are NEI results and the dashed lines are for the equilibrium ionization assumption.
    (c) Intensity ratios, AIA 131/94 and AIA 131/171 for both NEI and EI calculations. 
    }
    \label{fig:linetied_aia_y}
\end{figure}

Figure \ref{fig:linetied_aia_y}(a) shows the variation of count rates of the chosen high-temperature SDO/AIA bands (AIA 94 and 131, Figure \ref{fig:temp_respon}) along the reconnection outflow direction (also see white vertical dashed lines in Figure \ref{fig:linetied_aia}, left panels). 
In upward outflows ($y > 0$), the temperature gradually increases from $\sim 1.2 \times 10^7$ K to $\sim 1.8 \times 10^7$ K while the density slightly decreases as the plasma flows to a higher altitude from the reconnection X-point because the ambient coronal density decreases with the height due to the gravity (Figure \ref{fig:linetied_aia_y}(a)).
The AIA 94 and 131 count rates in EI dramatically decrease at the high-temperature end due to temperature (and partly density) variations, as shown in Figure \ref{fig:linetied_aia_y}(b).
The AIA 171 maintains relatively lower count rates because of the low-temperature response in high-temperature ranges (e.g., $\log(T) > 6.8$ K, in Figure \ref{fig:temp_respon}), and its variation along the $y-$ direction is more dominated by the density distribution.
In contrast, the AIA 131 intensity in NEI increases away from the reconnection X-point due to the under-ionization. In addition, AIA 94 and 171 count rates are all significantly enhanced in NEI calculations compared with the EI results.
The under-ionization of the current sheet also causes different features in different AIA intensity ratio profiles.  Figure \ref{fig:linetied_aia_y}(c) shows ratios of the brightest lines: AIA 94/131 and AIA 171/131. For comparison, we also plot these ratios in EI cases using dashed lines.
Because the NEI effect causes enhanced AIA 131 emissions in the reconnection upflow regions where the temperature increases with the height, the AIA 171/131 ratio (orange line) in NEI decreases as the reconnection current sheet temperature continually increases to $1.8 \times 10^7$ K, which is reversed from the AIA ratio (Figure \ref{fig:temp_respon}) based on EI. 
The AIA 94/131 ratio is also notably higher than the AIA 171/131 ratio due to NEI effects.

In downward outflow regions, the temperature is relatively lower compared with the upward outflows, and the density drops near the low tip of the reconnection current sheet.
%[A density drop would cause the count rate to go down.  Does this mean AIA94 compared to the other AIA bands?]
In equilibrium ionization assumptions, the AIA 94/131 and AIA 171/131 ratios are expected to increase at the low tip of the reconnection current sheet, when the temperature drops to $\sim 6 \times 10^6$ K.
However, the over-ionization greatly diminished this tendency as shown by red shadowing regions in Figure \ref{fig:linetied_aia_y}(c). The NEI AIA 94/131 (and AIA 171/131) ratios are then less than the EI assumption about one order of magnitude at $y \sim -0.6L_0$ accordingly.

%=======================================
\subsection{Radiative Cooling in NEI}
%=======================================
\begin{figure}
\plotone{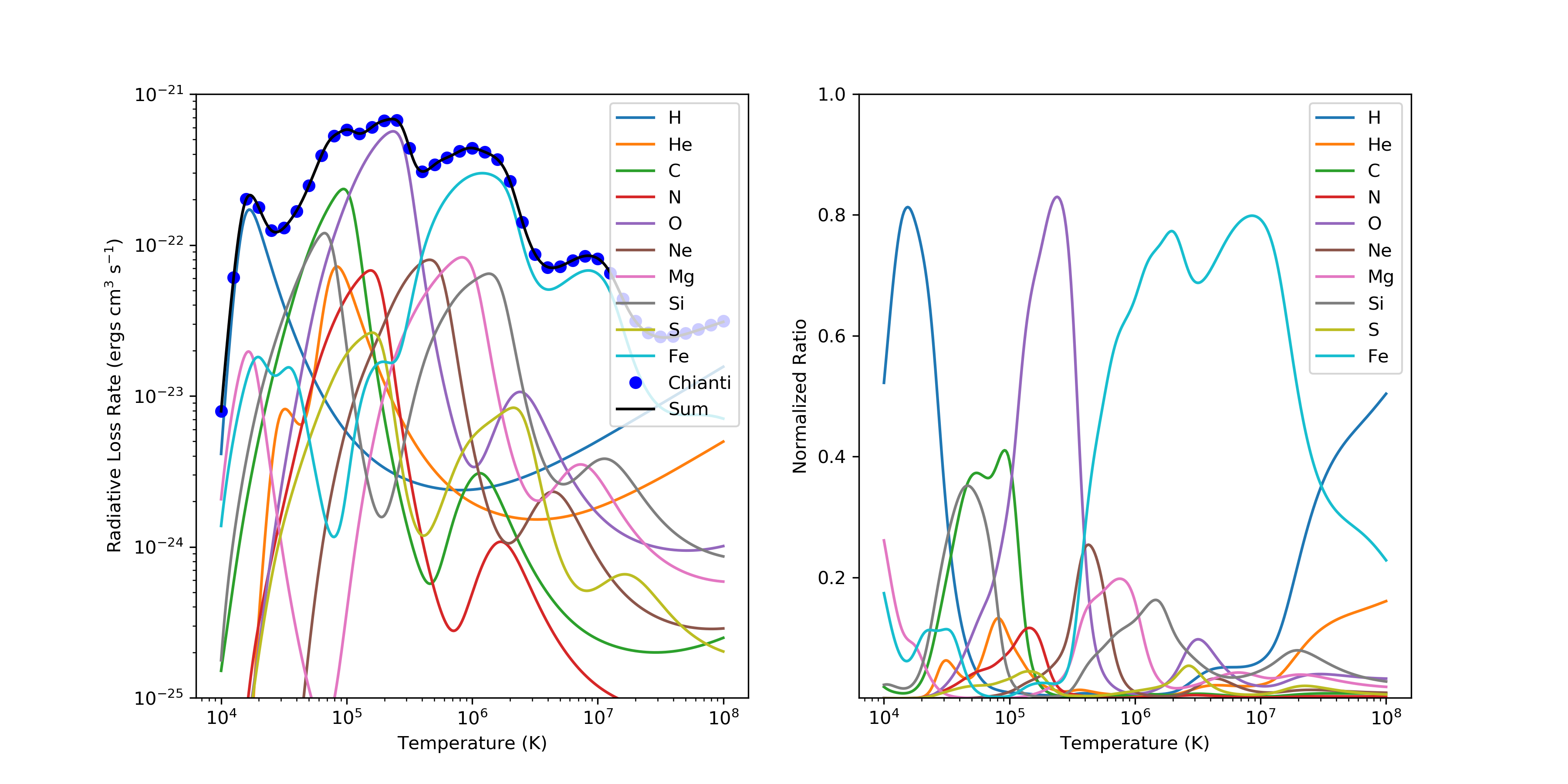}
\caption{Left panel: The radiative cooling rate of 10 abundant coronal elements saved in the table. The rates are calculated using the CHIANTI database (version 9). In this plot, 
we use the elemental abundance from the compilation by \cite{Schmelz2012ApJ...755...33S}
, and all ion fractions are assumed to be in equilibrium ionization.  Right panel: the relative contribution of each element on the total loss rate based on the same assumption as in the left figure.
\label{fig:cooling rate}}
\end{figure}

Optically thin radiative cooling can be affected by NEI when the ion population significantly departs from the equilibrium ionization assumption. Once the NEI population is known, the radiative loss rate can be directly calculated by summing over all the transitions for all abundant ions.
In solar coronal environments, the dominant radiative process includes bound-bound emission, while free-free, bound-free, and two-photon transitions also have significant contributions to high (and low) temperature plasma. So, we consider the above four transition processes in the following calculations.
Because the radiative emissions are functions of electron temperature but are less sensitive to density (aside from the proportionality to $\rho^2$), we neglect the density dependence and assume the electron density is $10^9$\ cm$^{-3}$ in the current analysis. However, it should be easy to include wider density ranges in the particular calculations.

In order to combine NEI cooling into the MHD simulation, the total energy density change due to optical thin radiative cooling can be described in the following form:
\begin{equation}
    \frac{\partial{E}}{\partial t} = -\rho^2 \Lambda(T, f_i).
\end{equation}
Here $E$ is the total energy density, $\rho$ is the plasma density, and $\Lambda(T, f_i)$ is the total radiative cooling rate which is the function of temperature ($T$) and ion fractions ($f_i$). 
%We obtain the rate $\Lambda_i(T, f_i)$ for each ion based on the CHIANTI atomic database \citep{Dere_2019ApJS..241...22D}, and save it into tables for the abundant elements in the corona as shown in Figure \ref{fig:cooling rate}. 
We calculate the cooling rate for each ion charge state, including the above four transition processes without ion fraction and abundance assumptions. Thus, the total cooling rate $\Lambda(T, f_i)$ can be updated with ion fraction either in NEI or EI as the form:
\begin{equation}
    \Lambda(T, f_i) = \displaystyle\sum_{elements} Abund \times (\displaystyle\sum_{charge\ states} f_{i} \times RadLoss_4(T)),
    \label{eq:cooling},
\end{equation}
here $Abund$ is the element abundance, $f_i$ is the ion fraction of the $i$th charge state based on NEI (or EI) calculations, and $RadLoss_4(T)$ is the cooling rate including free-free, bound-free, bound-bound and two-photon emissions obtained from Chianti database \citep{Dere_2019ApJS..241...22D} on a temperature grid. In cases where $f_i$ is set to be the equilibrium ionization fractions, the total $\Lambda(T, f_i)$ will be the same as the one in CHIANTI database (solid blue cycles in Figure \ref{fig:cooling rate}).

In combined MHD-NEI simulations, the radiative loss rate $\Lambda$ is computed by interpolating on temperature grids using local NEI ion fractions. It can be separated into the temperature-dependent cooling functions and the ion fractions, and be used in either one-step integration with time or multiple time-step schemes in an MHD solver. 
In the particular MHD-NEI simulation, one may not include all elements in the NEI module. So the cooling rate $\Lambda$ can be modified only for several abundant elements with the updated NEI fractions, and all other rates are approximated under the equilibrium ionization. For example, in the high-temperature coronal plasma ($10^6\mbox{~K} \sim 10^7\mbox{~K}$), the most critical elements for cooling loss are Fe, Si, Mg, and O, as shown in the right panel of Figure \ref{fig:cooling rate}.

\begin{figure}
    %% data: run_cs142_21Jun13b_eicool vs run_cs142_21Jun13b_neicool
    \plotone{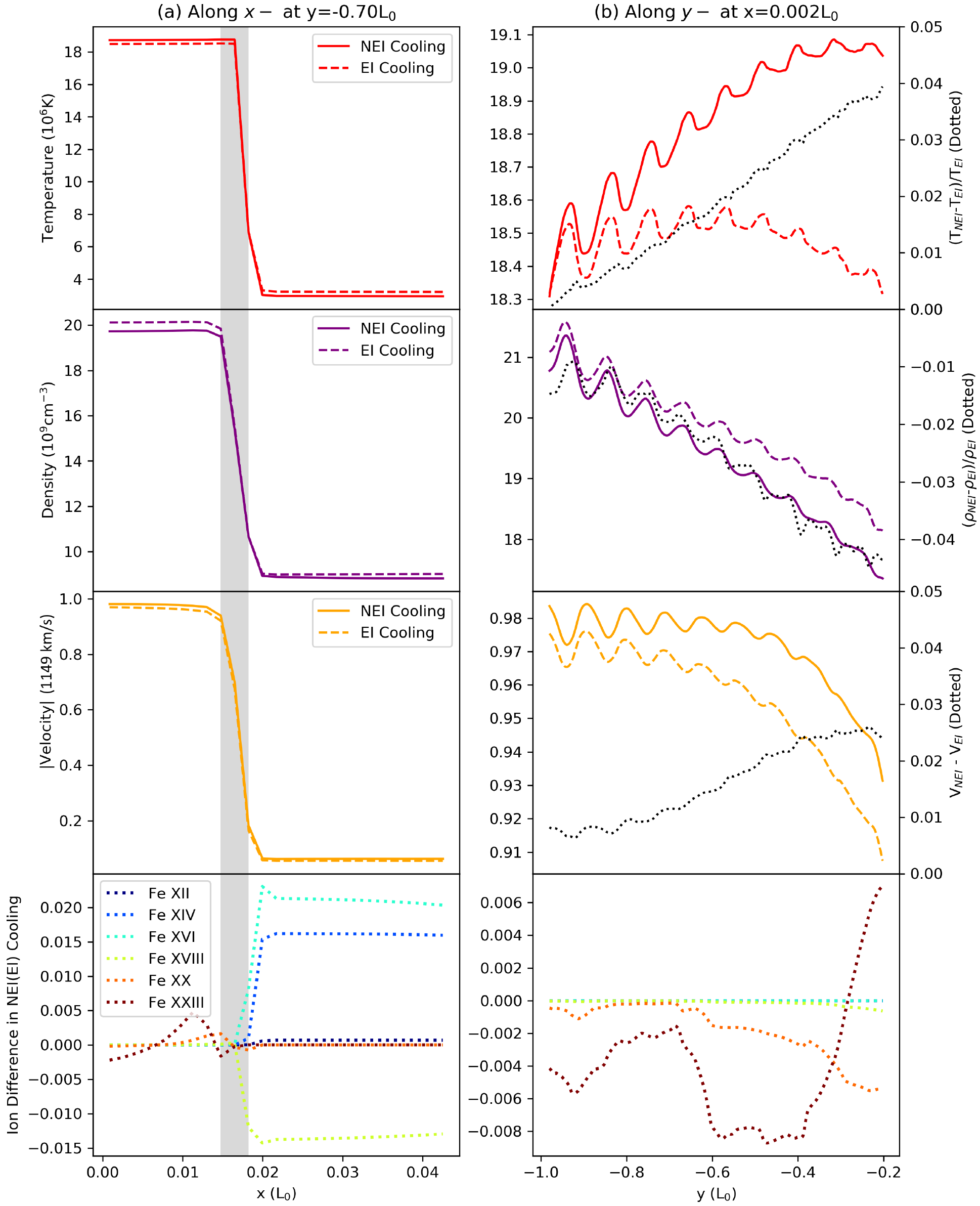}
    \caption{Comparison of primary variables and ion charge states between the NEI-dependent radiative cooling case (solid lines) and the EI-dependent cooling (dashed lines).
    (a) Temperature, density, and plasma flow velocity along reconnection inflow direction ($x-$) at $y=-0.7 L_0$. The vertical gray shadows indicate the position of the shock front, and the reconnection X-point is located at [$x=0, y=0$] as discussed in above Figure\ \ref{fig:13gh_halo_vs_nohalo}. The bottom panel shows Fe ion fraction differences defined by $f_{i,\mathrm{NEI~cooling}}-f_{i,\mathrm{EI~cooling}}$.
    (b) Similar to (a) but along the reconnection outflow direction ($y-$) at $x=0.002 L_0$. The right axis of each panel shows differences between the two cases for temperature, density, and velocity profiles, respectively.
    \label{fig:necooling_vs_eicooling}}
\end{figure}

As shown in Equation (\ref{eq:cooling}), the radiative cooling is proportional to $\rho^2$, so the cooling affects high-density plasma, where the NEI impact will be diminished.  In general, radiative cooling is expected to be smaller than expansion or thermal conduction cooling in high-temperature magnetic reconnection regions.  In addition, radiative cooling is very important in the cooler, denser prominence cores of CMEs, but the high density will limit the differences between EI and NEI.
However, due to the under-ionized nature inside Petschek-type reconnection current sheets, the systematic departure of cooling rates between NEI and EI in reconnection outflows may still affect the long-term evolution of the magnetic reconnection configuration.

Therefore, we compare two runs with different cooling rates using similar parameters based on Case A2.
In the two runs, the cooling term $-\rho^2 \Lambda(T, f_i)$ has been added into the energy equation, Equation (\ref{eq:energy}).
Because of radiative cooling terms, the temperature of background plasma could continually decrease to be much lower than the initial values. However, in order to obtain reasonable ionization states in the background corona, the upstream plasma should be in (or be close to) the typical corona temperature (e.g., 2MK).
Therefore, to balance the energy loss due to the radiative cooling, we included an artificial corona heating term (${\rho H_0}$) in the source term, ${S = \mu_0 \eta_{m}{j^2} + \nabla_{\|}\cdot\kappa\nabla_{\|} T + {\rho H_0} - \rho^2 \Lambda(T, f_i)}$. Here, $H_0$ is a constant heating rate defined by temperature so that $\rho H_0$ balances the radiative cooling term at the beginning. 
In Figure \ref{fig:necooling_vs_eicooling}, solid lines display the case employing the cooling rate from NEI results, and dashed lines are for cooling rates based on equilibrium ionization assumptions.
An overview feature is that NEI cooling rates cause a slightly hotter reconnection outflow. Inside the current sheet, the largest temperature difference between EI and NEI cooling is around $\sim$ 5\%. Meanwhile, the outflow speed in the NEI case is slower than in EI. However, the chosen Fe ion fraction is basically the same among the two runs, with a difference $f < \sim 2\%$.
We noticed that the primary variable profiles (including $T, \rho$, and $v$) show some rippling perturbations with height due to the numerical issues driven by the initial unbalance between heating and cooling once the magnetic reconnection occurs at the reconnection X-point at the beginning. However, these ripples are small, and we can still obtain the long-time tendency of NEI effects on reconnection configurations.
Thus, the NEI cooling effect is negligibly small in the analysis of Fe ion populations during short reconnection events (e.g., a few of Alfvén times). On the other hand, NEI cooling also contributes to temperature and density structures and should be further considered, especially in long-duration MHD simulations.

\section{Discussion}

%1.Multiple reconnection theories for heating plasma: shocks, turbulence, and multiple X-points -> affect the heating history -> under-ionized plasma mixed with over-ionized plasma in large scale reconnection sheet 
The magnetic reconnection is expected to affect the temperature distribution along the reconnection current sheet significantly. In reconnection theories, the competing heating mechanism include Petschek shock heating, multiple X-points and plasmoids, and turbulent heating, which may cause entirely different ionization features. 
In the Petschek shock configuration, the layered feature of ionization is reported in recent studies\citep[e.g.,][]{Imada_2011ApJ...742...70I, Imada2021ApJ...914L..28I}.
However, the heating due to multiple small-scale plasmoids and/or small-scale short-lived heating regions in turbulent reconnection \citep{Lazarian1999} may cause more complex ionization features.
In recent numerical modeling, several kinds of research suggest that those differences can simultaneously appear in large-scale reconnection current sheets \cite[e.g.,][]{Mei_2012MNRAS.425.2824M}. Therefore, high resolution MHD-NEI modeling is required to make meaningful comparison with the observations.

%2. Non-Maxwellian distribution around the shock -> ionization/recombination rates -> under-estimated ionization processes
As mentioned in the previous sections, a Maxwellian distribution of electrons is assumed in our current models. The ionization and recombination rates, as well as ionization states, are all based on this assumption. However, a significant fraction of electrons in the flare reconnection region can be accelerated into a power-law energy spectrum. Such energetic particles can significantly affect the ionization/recombination processes. For example, if the non-thermal electron distribution contains excess particles at about 5 to 10 times the mean energy, they can significantly increase ionization rates. The NEI effect may be enhanced around the shock front, where the electrons include non-thermal tails. Therefore, future studies should include both time-dependent ionization and non-thermal particles to get more accurate results for diagnostic studies of erupting plasma.

%3. Mixtures of ions -> 3D turbulent flows are expected and reported in previous models -> needs to be calculated in the near future
Our current model is based on single-fluid MHD simulations. In particular, the element abundance is assumed to be uniform in the whole simulation domain. However, in realistic large-scale models of solar eruptions, the reconnection inflows around the reconnection current sheet may come from  lower altitudes \citep[e.g.,][]{shen2013a}, where the plasma could already mix with chromosphere components. 
The background element abundance around the reconnection current sheet should be obtained by considering the whole plasma flowing history. Thus, the element abundances of the current sheet might evolve, affecting the optically thin radiative cooling calculations and the predicted EUV emission.
%\textbf{Furthermore, the different speeds of heavy ions are also reported in solar corona \citep[e.g.,][]{Gloeckler1995SSRv...71...79G}.} 
Therefore, a further study of the ionization calculation based on a multi-fluid model will be required.

% thermal conduction
The importance of thermal conduction in reconnecting current sheets has been commonly recognized in both analytic and numerical models \citep[e.g.,][]{Yokoyama1997ApJ...474L..61Y, Chen1999ApJ...513..516C, Yokoyama2001ApJ...549.1160Y,  Reeves_2010ApJ...721.1547R}. However, there are rare direct measurements of conduction rates around the reconnection current sheet during solar eruptions. 
In a 3D solar eruption model, \cite{Reeves2019ApJ...887..103R} performed the energy budget analysis around the post-CME reconnection current sheet, and predicted synthetic XRT and SDO/AIA images. They found that thermal conduction transports thermal energy away from the current sheet region, and widens the region of high temperatures.
On the other hand, the `thermal halo' may not be easily identified in realistic SDO/AIA observations due to instrumental limitations. \cite{Seaton2017ApJ...835..139S} performed differential emission measure analysis and revealed a highly uniform, hot current sheet with a cooler background. The results suggested that if the thermal halo is present, it is either too faint or too narrow a region to be detected in this eruption event. As discussed in these researches, the thermal halo could even be contained entirely within the reconnection current sheet regions or becomes important near the tip of the current sheet above the flare loop-tops.
In either case, the NEI analysis around the reconnection current sheet should serve as a necessary tool for understanding the nature of thermal energy transportation across the slow-mode shock front.

%4. the cost source and running cost
It is worth estimating the computational cost of the in-line NEI module during MHD simulations. As mentioned above,  we employ the passive scalar of the mass equation (Equation (\ref{eq:mass})) to store and update the ion population in the current calculation. Therefore, the required computer memory entirely depends on the number of ions to be solved. For instance, the MHD code saves at least 14 physical variables in memory for each cell (e.g., eight essential conserved variables, three interface magnetic fields, and three resistivity coefficients). In a particular model only including the Fe ions, the total number of ionization fractions occupying computer memory is around two times the original MHD variables. However, the cost may become more extensive, including all 14 abundant elements, which should be solved using multiple parallel cores and memories during simulations.
Our calculations also show that the run times of combined MHD-NEI modeling will be comparable with pure MHD simulations for the most abundant elements (e.g., C, O, or Fe). For instance, the model with 27 Fe ions spent $\sim$1.8 times the computer time compared with the pure MHD simulation with the same spatial resolution. In our cases, the massive runs, including all 14 elements, take about seven times longer than the Fe-only calculation. However, it is noticeable that the above run-time estimations are also significantly affected by the chosen MPI communication module on the cluster computer\footnote{Smithsonian High-Performance Cluster: https://doi.org/10.25572/SIHPC.}. Therefore, the run-time could be decreased in the future in a more high-efficiency communication environment (e.g., shared-memory machine). Thus, with only a modest increase in computing resources, NEI information will be conveniently achievable with available computing resources and allow a new accuracy level in diagnostic predictions.
The NEI module used in this work has been made available to the public and can be freely obtained from the Web\footnote{NEI module: https://doi.org/10.5281/zenodo.6555135.}. The eigenvalue and eigenvector tables associated with updated ionization/recombination rates, and cooling rates are also collected in an open-source program\footnote{NEI tables: https://github.com/ionizationcalc/time\_dependent\_fortran.}.

\section{Conclusion} \label{sec: conclusion}
Using the combined magnetohydrodynamics (MHD) and non-equilibrium ionization (NEI) simulations, we analyze the Petschek-type magnetic reconnection current sheet and ionization charge states in solar corona environments. Based on the NEI results, we also predict emission features observable by EUV instruments (SDO/AIA) for both ideal Petschek-type slow-mode shocks and solar flare reconnection current sheets.  

NEI is essential for making accurate, self-consistent predictions of emissions from high-temperature plasma during a solar eruption. In this work, we incorporated a robust NEI solver \citep{Shen2015} into the well-developed MHD code, Athena \citep{Stone2008ApJS..178..137S}, to perform in-line MHD-NEI simulations. The solver then can solve time-dependent ionization equations in each time-step during the MHD simulation and obtain information of evolving ion charge states on the whole simulation domain. We employ this method in several test projects, such as shock tubes (see Appendix A). Comparison with post-processed NEI calculations shows that this in-line NEI module calculates charge states accurately, and it can be used in complex problems such as shocks and magnetic reconnection outflows.  
We then employed the above method in the Petschek-type magnetic reconnection configuration. We obtained the NEI properties for both the classical Spitzer thermal conduction and conductive flux-limited models. The features of the NEI and EUV emission around the reconnection current sheet are studied and summarized:

1. The high-temperature thermal halo around the Petschek shock front due to thermal conduction can be found in the classical Spitzer conduction model. The width of the halo can be reduced to the simulation grid size in conductive flux-limited models. In both cases, the NEI significantly affects ion population distributions and causes under-ionized features inside the reconnection current sheet as well as the thermal halo region. 

2. In an ideal Petschek reconnection current sheet, the two-dimensional spatial distribution of ion charge states can be estimated by analyzing the 1D plasma evolution along the outflow direction once the shock angle is known, assuming that the current sheet is uniform. However, the MHD-NEI simulations will be necessary to get accurate ionization calculations because of the temperature variation in the reconnection outflow direction \citep{ko2010}.

3. A dense reconnection current sheet generally reduces NEI effects, especially for $\rho > \sim 10^{10}$ cm$^{-3}$. On the other hand, NEI effects also substantially depend on the temperature of observational targets: the higher charge states (e.g., \ion{Fe}{20} - \ion{Fe}{25}) depart more from the equilibrium ionization (EI) results compared with the lower charge states (e.g., \ion{Fe}{14} - \ion{Fe}{18}). This occurs because equilibration timescales for the higher ionization ions are substantially longer than those of the low ionization ions \citep{Smith2010ApJ...718..583S}.  
Our results indicate that the dominant ions in extremely high-temperature reconnection current sheets (e.g., 16 MK and above) should be obtained by solving time-dependent ionization equations, while the most abundant ions in a relatively cool current sheet (e.g., 8MK) are barely affected by NEI.

4. Synthetic EUV narrowband SDO/AIA images are compared between EI and NEI predictions.
Inside the reconnection current sheet, the EI assumption under-estimated AIA 94, 131, 171, and 335 and over-estimated AIA 193 in the typical high-temperature plasma ($\sim$ 16 MK). Around the thermal halo regions, the variation of all AIA count rates substantially depends on the width of the halo region. This model suggests that the gap between the different AIA intensity gradient (such as AIA 94, 131, 171, and 211) around the shock front could serve as a possible diagnostic of shocks and halos. 

We estimated the departure of the NEI from the EI assumption on the reconstructed temperatures using multiple SDO/AIA band images in typical hot reconnection current sheets ($\sim 10^7$ K) and extremely high-temperature environments ($\sim$ 16 MK). 
The under-ionized nature inside the reconnection current sheet causes a significant underestimation of the EM reconstructed temperature. Near the current sheet edge, the NEI reconstructed temperature can be lower by $\sim$ 60\% compared to the actual temperature. The results show that the apparent width of the hot current sheet based on the emission reconstruction method may decrease by $\sim$ 8\% and 31\% in different density situations ($\sim 10^{10}$ and $\sim 2.5 \times 10^9$ cm$^{-3}$).

In a vertical reconnection current sheet configuration formed during solar flare eruptions, the bi-directional reconnection outflows have been revealed in our models. 
The upward reconnection outflows are under-ionized, similar to the Petschek-type reconnection plasma. The downward reconnection outflows appear to have a slower speed due to the line-tied boundary condition and the formation of flare loops. 
The temperature and density at the lower tip of the reconnection current sheet quickly drop, which may cause over-ionized features, opposite to the classical Petschek-type reconnection flows.
In tenuous plasma situations (e.g., $\sim 2.5 \times 10^9$ cm$^{-3}$), the corresponding synthetic EI intensity ratios of multiple bands (AIA 94/131, AIA 171/131), therefore, could be significantly over-estimated by about one order of magnitude comparing the NEI predictions.
%[Since NEI is ocrrect, it is better to say that EI overestimates the ratios.]

%\begin{figure}
%\begin{interactive}{animation}{movie.mp4}
%\plotone{f4.pdf}
%\end{interactive}
%\caption{Figure 1 from \citet{2018ApJ...868L..33L}. AIA 171\AA (a,b), 
%IA 131\AA (c), and AIA 304\AA images are shown. The red rectangle 
%in (a) shows the field of view of the other panels. An animation of 
%panels (b-d) is available. It covers 8 hours of observing beginning 
%t 01:00 UT on 2012 January 19. The video duration is 20 seconds. 
%\label{fig:video}}
%\end{figure}

%Animations and interactive figures (Section \ref{sec:interactive}) should 
%use the environment in the figure call. This
%environment
%places a blue border around the figure to indicate that the figure is 
%enhanced in the published HTML article. The
%command also serves to alert the publisher what files are used to generate
%the dynamic HTML content. {\tt\string\interactive} takes two arguments. The
%first details the type and currently only three are allowed. The types are
%{\tt\string js} for generic javascript interactive figures, 
%{\tt\string animation} for inline videos, and 
%{\tt\string timeseries} for interactive light curves produced
%by astropy \citet{2013A&A...558A..33A}\footnote{To be release in the 
%summer of 2019}. If these types are not provide the compiler will issue an
%error and quit. The second argument is the file that produces the enhanced
%feature in the HTML article.

\acknowledgments
The authors thank Jing Ye for suggestions that helped to perform MHD simulations, and thank the anonymous reviewers for valuable comments to improve the paper. This work was supported by NSF grants AGS-1723313 and AST 2108438 to the Smithsonian Astrophysical Observatory.
The work also supported by NASA grants 80NSSC21K2044, 80NSSC19K0853, 80NSSC18K1129, and 80NSSC20K1318.
CHIANTI is a collaborative project involving George Mason University, the University of Michigan (USA), and the University of Cambridge (UK). The computations in this paper were conducted on the Smithsonian High Performance Cluster, Smithsonian Institution (https://doi.org/10.25572/SIHPC).


%\vspace{5mm}
%\facilities{HST(STIS), Swift(XRT and UVOT), AAVSO, CTIO:1.3m,


\software{Athena \citep{Stone2008ApJS..178..137S}, 
          NEI solver \citep{Shen2015zndo....272609S},
          Chiantipy \citep{chiantipy2013ascl.soft08017D},
          astropy \citep{2013A&A...558A..33A},
          sunpy \citep{sunpy2022zndo....591887M}
          }



\appendix
\begin{figure}
\plotone{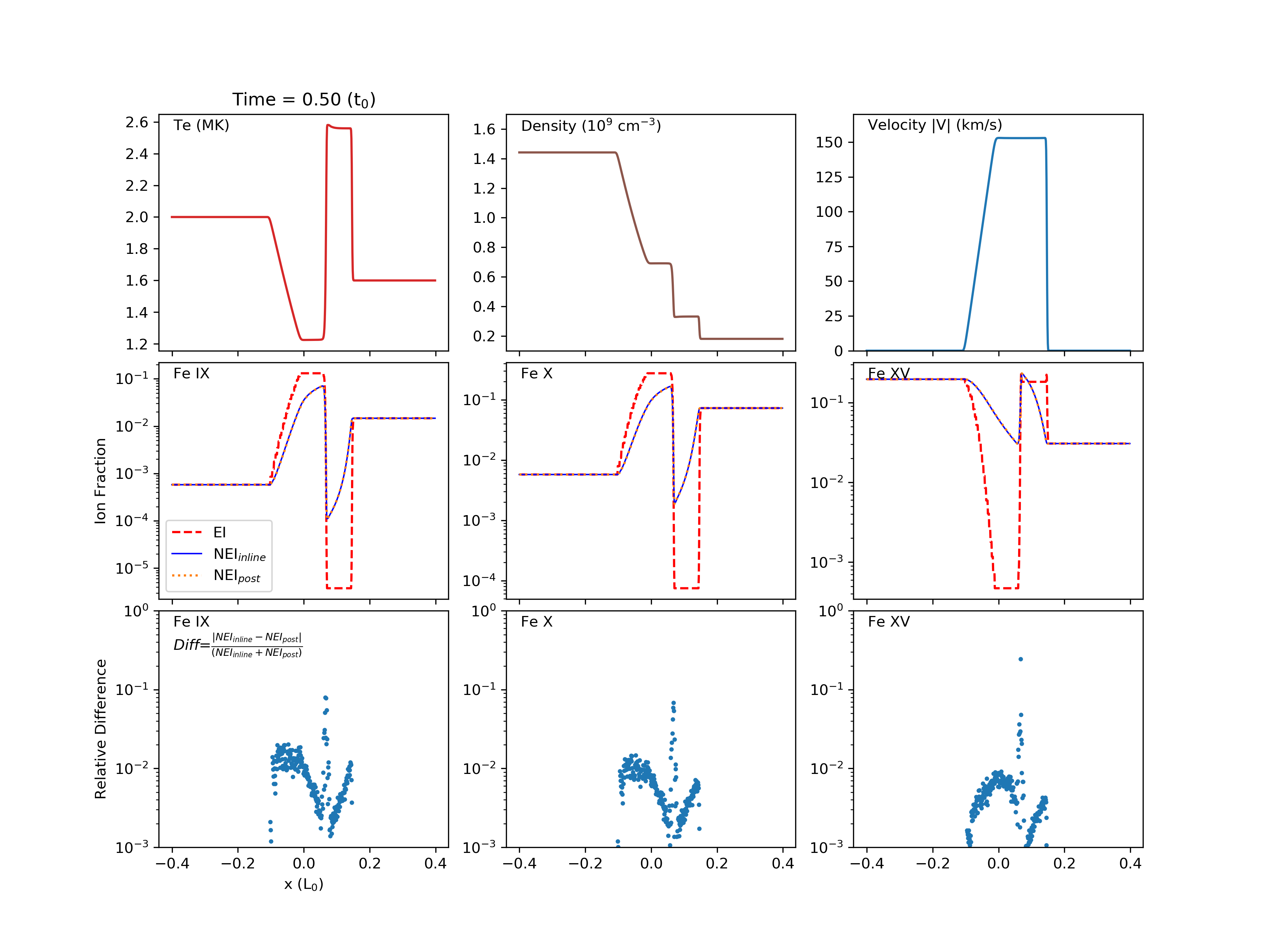}
\caption{The comparison of the chosen ion fractions between in-line NEI simulations and the post-processed NEI calculation at the sampling points in 1D Sod shock test. The second row shows ion fractions of Fe IX, Fe X, and Fe XIV, and the third row is the relative difference between in-line NEI and post-processed NEI fractions.
\label{fig:1d_diff}}
\end{figure}

\begin{figure}
    % data: run_cs142_21May22f
    \plotone{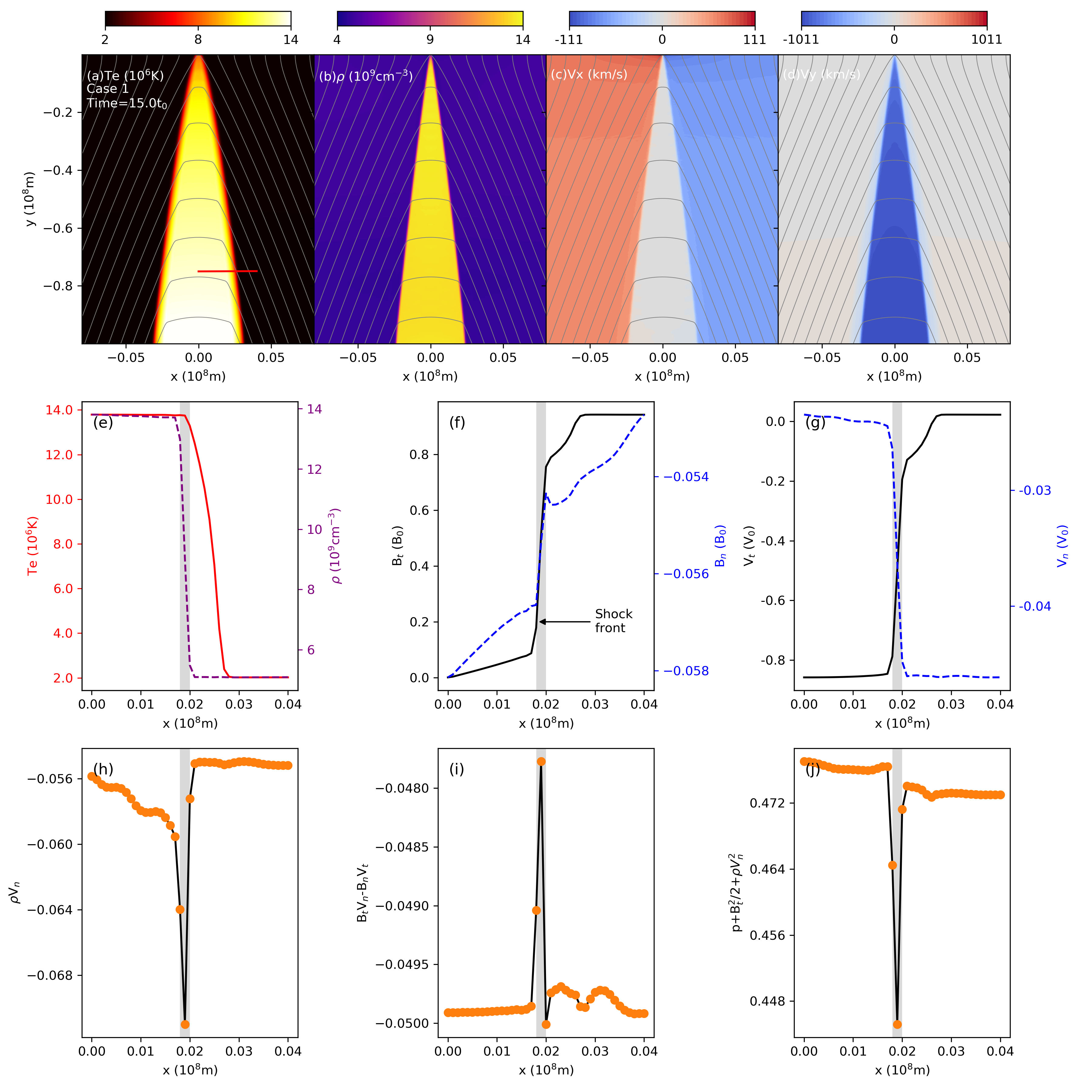}
    \caption{The slow-mode shocks in a Petschek-type reconnection simulation with the classical Spitzer thermal conduction coefficients. (a-d)
    The temperature, density, plasma velocity around the reconnection X-point, and a pair of slow-mode shocks. The solid gray contours are for magnetic field lines.
    (e-g) Primary variables distribution crossing the shock front along the red sampling line shown in the panel (a). The vertical gray shadows indicate the position of the shock front. Here $B_t$ (and $V_t$) is the transverse component of magnetic field (and velocity), and $B_n$ (and $V_n$) is the normal component to the shock front, respectively. (h-j) Non-dimensional mass flux, magnetic flux, and energy flux distribution that continues crossing the slow mode shocks.
    \label{fig:sk_thermalhalo}}
\end{figure}

\begin{figure}
    \centering
    \plotone{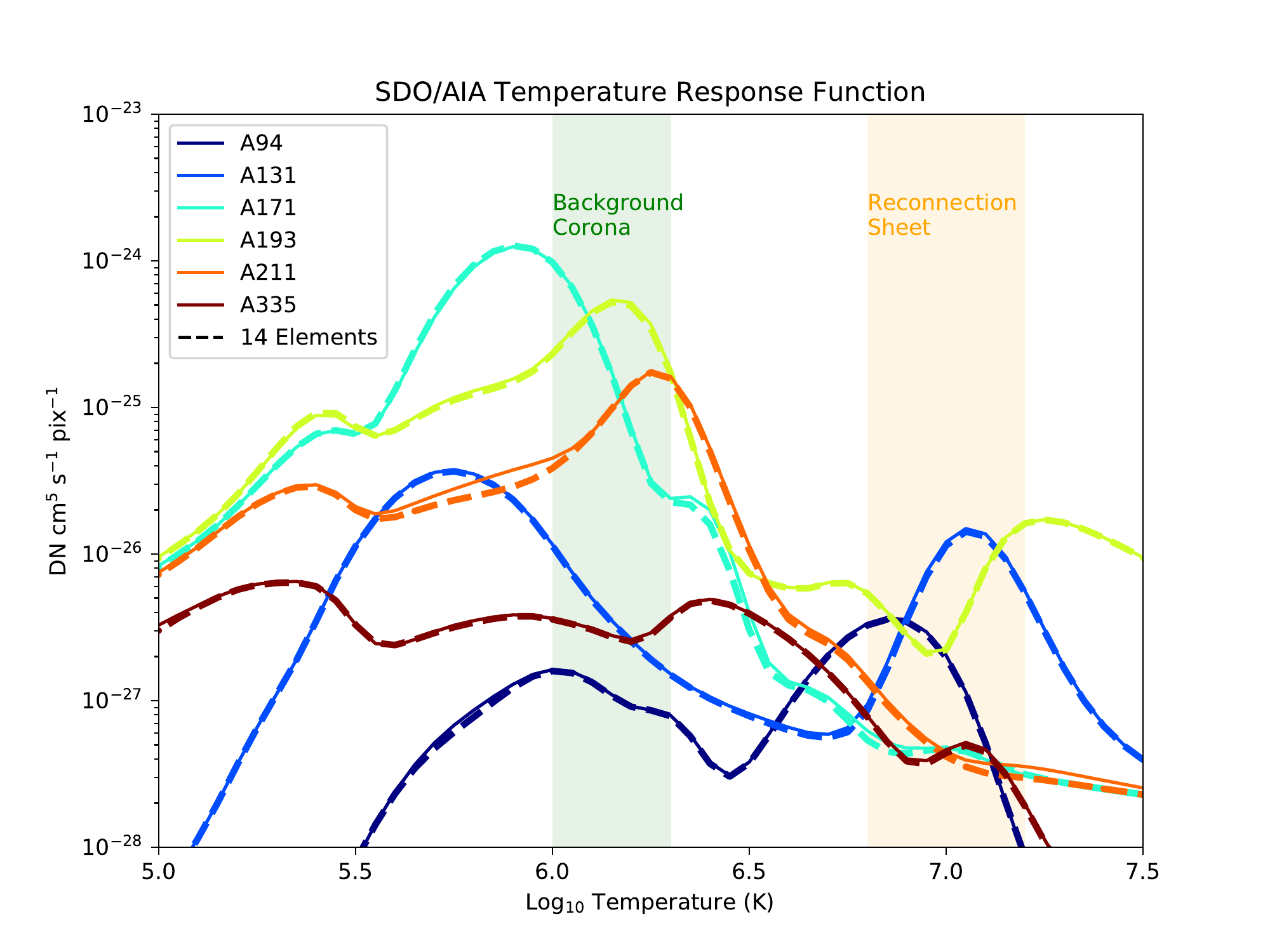}
    \caption{Temperature response function for six SDO/AIA channels. The solid lines are calculated from the Chianti database \citep{Dere_2019ApJS..241...22D} assuming ionization equilibrium and the coronal abundance \citep{Feldman1992PhyS...46..202F}. The dotted lines counts the most abundant 14 elements (H, He, C, N, O, Ne, Mg, Al, Si, S, Ar, Ca, Fe, and Ni) calculated in this work. The green and orange shaded regions indicate the temperature ranges of the background corona and reconnection current sheets, respectively.}
    \label{fig:temp_respon}
\end{figure}

\section{1D shock-tube problem}
% 1D shock front
The shock=tube problem has been commonly used in MHD simulation as a standard test. Here we perform the 1D shock=tube simulation based on the classical Sod shock=tube configuration where two constant states have been separated by a discontinuity at the beginning. We apply the solar coronal temperature ($10^6$ to $10^7$ K) in this test and setup the non-dimensional plasma density and gas pressure as the following: ${\rho_{l} = 1, p_{l}=0.025}$, and ${\rho_{r}=0.125, p_{r}=0.0025}$. Here the subscript $l$ and ${r}$ indicate the left and right hand side of x=0 along the $x$- axis, respectively. 
Figure \ref{fig:1d_diff} shows the results for the temperature, density, velocity, and several Fe ion fractions at $t=0.5t_0$) on the grid of 1000 cells in the domain $-0.5 < x < 0.5$. In the second row of Fig.\ \ref{fig:1d_diff}, the \ion{Fe}{9}, \ion{Fe}{10}, \ion{Fe}{15} ion fraction distributions as functions of $x$ are plotted. The red dashed lines are for ionization equilibrium profiles which strongly depend on plasma temperature, while blue lines are for in-line NEI results. For comparison,  we also perform the post-processed NEI calculation (see red dotted lines) by tracing the trajectory of plasma movements and solve the time-dependent ionization equation in the Lagrangian frameworks to get ion fractions \citep[e.g.,][]{shen2013b,Shen2015}. 
In the third row, we checked the difference between in-line NEI results and post-processed NEI calculations by plotting the $Diff = \frac{NEI_{inline} - NEI_{post}}{NEI_{inline}+NEI_{post}}$. 
It is clear that the errors are lower than about two percent for the above-chosen ions in postshock regions. $Diff$ could be larger at a few points around the shocks caused by the interpolation process (both spatially and temporally) while tracing the movement of plasma in the post-processed NEI calculations. Therefore, we expect the error can be minor with fine simulation cells and outputting intervals.

% Jump-conditions
\section{Jump condition}
In order to confirm the shock properties in detail, we check the primary variables across the shock front along the red solid line in Figure \ref{fig:sk_thermalhalo}(a)). This sampling line is chosen to be perpendicular to the shock surface, and has an angle of $1.35^\circ$ with $x-$ direction. Figure \ref{fig:sk_thermalhalo}(e-g) shows that these variables jump from the shock upstream (right side along $x-$ direction) to the downstream (left side), including temperature, density, magnetic, and velocity field. The shock front is indicated by the vertical gray shadows. It is interesting that a "thermal halo" region, outside the high-density reconnection current sheet (see Figure \ref{fig:sk_thermalhalo}(a) and (e)), also can be clearly seen due to the thermal conduction with the classical Spitzer thermal conduction coefficient in this simulation. However, this "thermal halo" could be over-estimated in the model, and has been discussed in above sections. In the third row of Figure \ref{fig:sk_thermalhalo}, we further calculated the mass flux component, magnetic flux component, and momentum flux along the cut line. As can be seen in Figure \ref{fig:sk_thermalhalo}(h, i), the relative changes for $\rho V_n$ and $B_t V_n - B_n V_t$ are roughly 6\% and less than $\sim$2\% between the upstream and downstream of the shocks, respectively. The momentum
flux component is also conserved, and the relative changes between upstream and downstream are still less than $\sim$1\% as shown in panel (j). 
It is noticed that there are unavoidable measurement errors when estimating the direction of the shock front, which may cause slight deviations to both interpolated transverse and normal components of primary variables. However, conservation conditions of primary variables are basically satisfied with the Rankine–Hugoniot relation between upstream and downstream.

%\section{2D blast waves}

%\begin{figure}
% data: workspaces/mhd_simulation/Athena/run4.2_nei/blast_nei_2d/vtk_20200426_120836
%\plotone{fig_2d_blast_initial.png}
%\caption{The temperature distribution at the beginning, 0.25$t_0$, and 0.5$t_0$ in the 2D blast wave test. The %solid lines indicate magnetic lines.
%\label{fig:2d_blast_initial}}
%\end{figure}

%\begin{figure}
%\plotone{fig_2d_blast_contour.png}
%\caption{The temperature, density, and velocity distribution at 0.5$t_0$ in the 2D blast wave test. The dotted line on temperature contour shows the sampling points we perform the post-NEI analysis. The second row shows ion fraction of Fe XII, Fe XVIII, and Fe XXI from the in-line NEI simulations. The third row is for equilibrium fractions.
%\label{fig:2d_blast_contour}}
%\end{figure}

%\begin{figure}
%\plotone{fig_2d_blast_line.png}
%\caption{The comparison of the chosen ion fractions between in-line NEI simulations and the post-NEI calculation on the sampling points . The second row shows ion fraction of Fe XII, Fe XVIII, and Fe XXI along a sampling line (gray dotted line in Fig. \ref{fig:2d_blast_contour}), and  the third row is the relative difference between in-line NEI and post-NEI fractions.
%\label{fig:2d_blast_line}}
%\end{figure}

% ------------------------------------------------------------------
% Temperature response function of SDO/AIA
% ------------------------------------------------------------------
\section{EUV Emission} 
Figure \ref{fig:temp_respon} shows the temperature response function for six SDO/AIA channels. The response function with 14 of the most abundant elements can be a good approximation in most coronal temperature ranges.


\bibliography{cs_nei}{}

\begin{thebibliography}{}
\expandafter\ifx\csname natexlab\endcsname\relax\def\natexlab#1{#1}\fi
\providecommand{\url}[1]{\href{#1}{#1}}
\providecommand{\dodoi}[1]{doi:~\href{http://doi.org/#1}{\nolinkurl{#1}}}
\providecommand{\doeprint}[1]{\href{http://ascl.net/#1}{\nolinkurl{http://ascl.net/#1}}}
\providecommand{\doarXiv}[1]{\href{https://arxiv.org/abs/#1}{\nolinkurl{https://arxiv.org/abs/#1}}}

\bibitem[{{Astropy Collaboration} {et~al.}(2013){Astropy Collaboration},
  {Robitaille}, {Tollerud}, {Greenfield}, {Droettboom}, {Bray}, {Aldcroft},
  {Davis}, {Ginsburg}, {Price-Whelan}, {Kerzendorf}, {Conley}, {Crighton},
  {Barbary}, {Muna}, {Ferguson}, {Grollier}, {Parikh}, {Nair}, {Unther},
  {Deil}, {Woillez}, {Conseil}, {Kramer}, {Turner}, {Singer}, {Fox}, {Weaver},
  {Zabalza}, {Edwards}, {Azalee Bostroem}, {Burke}, {Casey}, {Crawford},
  {Dencheva}, {Ely}, {Jenness}, {Labrie}, {Lim}, {Pierfederici}, {Pontzen},
  {Ptak}, {Refsdal}, {Servillat}, \& {Streicher}}]{2013A&A...558A..33A}
{Astropy Collaboration}, {Robitaille}, T.~P., {Tollerud}, E.~J., {et~al.} 2013,
  \aap, 558, A33, \dodoi{10.1051/0004-6361/201322068}

\bibitem[{{Bhattacharjee} {et~al.}(2009){Bhattacharjee}, {Huang}, {Yang}, \&
  {Rogers}}]{Bhattacharjee_2009PhPl...16k2102B}
{Bhattacharjee}, A., {Huang}, Y.-M., {Yang}, H., \& {Rogers}, B. 2009, Physics
  of Plasmas, 16, 112102, \dodoi{10.1063/1.3264103}

\bibitem[{{Carmichael}(1964)}]{Carmichael1964}
{Carmichael}, H. 1964, NASA Special Publication, 50, 451

\bibitem[{{Chen} {et~al.}(1999){Chen}, {Fang}, {Tang}, \&
  {Ding}}]{Chen1999ApJ...513..516C}
{Chen}, P.~F., {Fang}, C., {Tang}, Y.~H., \& {Ding}, M.~D. 1999, \apj, 513,
  516, \dodoi{10.1086/306823}

\bibitem[{{Cheng} {et~al.}(2018){Cheng}, {Li}, {Wan}, {Ding}, {Chen}, {Zhang},
  \& {Liu}}]{Cheng2018ApJ...866...64C}
{Cheng}, X., {Li}, Y., {Wan}, L.~F., {et~al.} 2018, \apj, 866, 64,
  \dodoi{10.3847/1538-4357/aadd16}

\bibitem[{{Cheung} {et~al.}(2015){Cheung}, {Boerner}, {Schrijver}, {Testa},
  {Chen}, {Peter}, \& {Malanushenko}}]{Cheung2015ApJ...807..143C}
{Cheung}, M. C.~M., {Boerner}, P., {Schrijver}, C.~J., {et~al.} 2015, \apj,
  807, 143, \dodoi{10.1088/0004-637X/807/2/143}

\bibitem[{{Ciaravella} \& {Raymond}(2008)}]{ciaravella2008}
{Ciaravella}, A., \& {Raymond}, J.~C. 2008, ApJ, 686, 1372,
  \dodoi{10.1086/590655}

\bibitem[{{Ciaravella} {et~al.}(2002){Ciaravella}, {Raymond}, {Li}, {Reiser},
  {Gardner}, {Ko}, \& {Fineschi}}]{Ciaravella2002ApJ...575.1116C}
{Ciaravella}, A., {Raymond}, J.~C., {Li}, J., {et~al.} 2002, \apj, 575, 1116,
  \dodoi{10.1086/341473}

\bibitem[{Cowie \& McKee(1977)}]{cowie_evaporation_1977}
Cowie, L.~L., \& McKee, C.~F. 1977, The Astrophysical Journal, 211, 135,
  \dodoi{10.1086/154911}

\bibitem[{{Dere}(2013)}]{chiantipy2013ascl.soft08017D}
{Dere}, K. 2013, {ChiantiPy: Python package for the CHIANTI atomic database},
  Astrophysics Source Code Library, record ascl:1308.017.
\newblock \doeprint{1308.017}

\bibitem[{{Dere} {et~al.}(2019){Dere}, {Del Zanna}, {Young}, {Landi}, \&
  {Sutherland}}]{Dere_2019ApJS..241...22D}
{Dere}, K.~P., {Del Zanna}, G., {Young}, P.~R., {Landi}, E., \& {Sutherland},
  R.~S. 2019, \apjs, 241, 22, \dodoi{10.3847/1538-4365/ab05cf}

\bibitem[{{Feldman}(1992)}]{Feldman1992PhyS...46..202F}
{Feldman}, U. 1992, \physscr, 46, 202, \dodoi{10.1088/0031-8949/46/3/002}

\bibitem[{{Forbes}(1986)}]{Forbes1986ApJ...305..553F}
{Forbes}, T.~G. 1986, \apj, 305, 553, \dodoi{10.1086/164268}

\bibitem[{{Forbes} \& {Priest}(1987)}]{Forbes1987RvGeo..25.1583F}
{Forbes}, T.~G., \& {Priest}, E.~R. 1987, Reviews of Geophysics, 25, 1583,
  \dodoi{10.1029/RG025i008p01583}

\bibitem[{{Forbes} {et~al.}(2018){Forbes}, {Seaton}, \&
  {Reeves}}]{Forbes_2018ApJ...858...70F}
{Forbes}, T.~G., {Seaton}, D.~B., \& {Reeves}, K.~K. 2018, \apj, 858, 70,
  \dodoi{10.3847/1538-4357/aabad4}

\bibitem[{Foster {et~al.}(2018)Foster, Smith, Brickhouse, Mullen, Cumbee,
  Stancil, \& Cui}]{Foster_atomdb_2018}
Foster, A., Smith, R., Brickhouse, N.~S., {et~al.} 2018, American Astronomical
  Society Meeting Abstracts \#231, 231, 253.03.
\newblock \url{https://ui.adsabs.harvard.edu/abs/2018AAS...23125303F/abstract}

\bibitem[{{Fryxell} {et~al.}(2000){Fryxell}, {Olson}, {Ricker}, {Timmes},
  {Zingale}, {Lamb}, {MacNeice}, {Rosner}, {Truran}, \&
  {Tufo}}]{Fryxell2000ApJS..131..273F}
{Fryxell}, B., {Olson}, K., {Ricker}, P., {et~al.} 2000, \apjs, 131, 273,
  \dodoi{10.1086/317361}

\bibitem[{{Hirayama}(1974)}]{Hirayama1974}
{Hirayama}, T. 1974, \solphys, 34, 323, \dodoi{10.1007/BF00153671}

\bibitem[{{Huang} \& {Bhattacharjee}(2016)}]{huang2016}
{Huang}, Y.-M., \& {Bhattacharjee}, A. 2016, ApJ, 818, 20,
  \dodoi{10.3847/0004-637X/818/1/20}

\bibitem[{{Hughes} \& {Helfand}(1985)}]{hugheshelfand1985}
{Hughes}, J.~P., \& {Helfand}, D.~J. 1985, ApJ, 291, 544,
  \dodoi{10.1086/163095}

\bibitem[{{Imada}(2021)}]{Imada2021ApJ...914L..28I}
{Imada}, S. 2021, \apjl, 914, L28, \dodoi{10.3847/2041-8213/ac063c}

\bibitem[{{Imada} {et~al.}(2011){Imada}, {Murakami}, {Watanabe}, {Hara}, \&
  {Shimizu}}]{Imada_2011ApJ...742...70I}
{Imada}, S., {Murakami}, I., {Watanabe}, T., {Hara}, H., \& {Shimizu}, T. 2011,
  \apj, 742, 70, \dodoi{10.1088/0004-637X/742/2/70}

\bibitem[{{Ko} {et~al.}(2003){Ko}, {Raymond}, {Lin}, {Lawrence}, {Li}, \&
  {Fludra}}]{ko2003}
{Ko}, Y.-K., {Raymond}, J.~C., {Lin}, J., {et~al.} 2003, ApJ, 594, 1068,
  \dodoi{10.1086/376982}

\bibitem[{{Ko} {et~al.}(2010){Ko}, {Raymond}, {Vr{\v s}nak}, \&
  {Vuji{\'c}}}]{ko2010}
{Ko}, Y.-K., {Raymond}, J.~C., {Vr{\v s}nak}, B., \& {Vuji{\'c}}, E. 2010, ApJ,
  722, 625, \dodoi{10.1088/0004-637X/722/1/625}

\bibitem[{{Kopp} \& {Pneuman}(1976)}]{Kopp-Pneuman1976}
{Kopp}, R.~A., \& {Pneuman}, G.~W. 1976, \solphys, 50, 85,
  \dodoi{10.1007/BF00206193}

\bibitem[{{Landi} {et~al.}(2012){Landi}, {Alexander}, {Gruesbeck}, {Gilbert},
  {Lepri}, {Manchester}, \& {Zurbuchen}}]{Landi2012ApJ...744..100L}
{Landi}, E., {Alexander}, R.~L., {Gruesbeck}, J.~R., {et~al.} 2012, \apj, 744,
  100, \dodoi{10.1088/0004-637X/744/2/100}

\bibitem[{{Landi} \& {Lepri}(2015)}]{Landi2015ApJ...812L..28L}
{Landi}, E., \& {Lepri}, S.~T. 2015, \apjl, 812, L28,
  \dodoi{10.1088/2041-8205/812/2/L28}

\bibitem[{{Lazarian} \&
  {Vishniac}(1999{\natexlab{a}})}]{Lazarian1999ApJ...517..700L}
{Lazarian}, A., \& {Vishniac}, E.~T. 1999{\natexlab{a}}, \apj, 517, 700,
  \dodoi{10.1086/307233}

\bibitem[{{Lazarian} \& {Vishniac}(1999{\natexlab{b}})}]{Lazarian1999}
---. 1999{\natexlab{b}}, ApJ, 517, 700, \dodoi{10.1086/307233}

\bibitem[{{Lee} {et~al.}(2019){Lee}, {Raymond}, {Reeves}, {Shen}, {Moon}, \&
  {Kim}}]{Lee2019ApJ...879..111L}
{Lee}, J.-Y., {Raymond}, J.~C., {Reeves}, K.~K., {et~al.} 2019, \apj, 879, 111,
  \dodoi{10.3847/1538-4357/ab24bb}

\bibitem[{{Lin} {et~al.}(2021){Lin}, {Liu}, \& {Li}}]{Lin2021PhPl...28g2109L}
{Lin}, S.-C., {Liu}, Y.-H., \& {Li}, X. 2021, Physics of Plasmas, 28, 072109,
  \dodoi{10.1063/5.0052317}

\bibitem[{{Lionello} {et~al.}(2019){Lionello}, {Downs}, {Linker}, {Miki{\'c}},
  {Raymond}, {Shen}, \& {Velli}}]{Lionello2019SoPh..294...13L}
{Lionello}, R., {Downs}, C., {Linker}, J.~A., {et~al.} 2019, \solphys, 294, 13,
  \dodoi{10.1007/s11207-019-1401-2}

\bibitem[{{Loureiro} {et~al.}(2007){Loureiro}, {Schekochihin}, \&
  {Cowley}}]{Loureiro_2007PhPl...14j0703L}
{Loureiro}, N.~F., {Schekochihin}, A.~A., \& {Cowley}, S.~C. 2007, Physics of
  Plasmas, 14, 100703, \dodoi{10.1063/1.2783986}

\bibitem[{{Masai}(1984)}]{masai1984}
{Masai}, K. 1984, Astrophysics and Space Science, 98, 367,
  \dodoi{10.1007/BF00651415}

\bibitem[{{Mei} {et~al.}(2012){Mei}, {Shen}, {Wu}, {Lin}, {Murphy}, \&
  {Roussev}}]{Mei_2012MNRAS.425.2824M}
{Mei}, Z., {Shen}, C., {Wu}, N., {et~al.} 2012, \mnras, 425, 2824,
  \dodoi{10.1111/j.1365-2966.2012.21625.x}

\bibitem[{{Mumford} {et~al.}(2022){Mumford}, {Freij}, {Christe}, {Ireland},
  {Mayer}, {Stansby}, {Shih}, {Hughitt}, {Ryan}, {Liedtke},
  {P{\'e}rez-Su{\'a}rez}, {Vishnunarayan K}, {Hayes}, {Chakraborty}, {Inglis},
  {Pattnaik}, {Sip{\H{o}}cz}, {Sharma}, {Leonard}, {Hewett}, {Barnes},
  {Hamilton}, {Manhas}, {Panda}, {Earnshaw}, {Choudhary}, {Kumar}, {Singh},
  {Chanda}, {Akramul Haque}, {Kirk}, {Mueller}, {Konge}, {Srivastava}, {Jain},
  {Bennett}, {Baruah}, {Arbolante}, {Maloney}, {Charlton}, {Mishra}, {Paul},
  {MacBride}, {Chorley}, {Himanshu}, {Chouhan}, {Modi}, {Sharma}, {Mason},
  {Naman9639}, {Zivadinovic}, {Bobra}, {Campos Rozo}, {Manley}, {Ivashkiv},
  {Chatterjee}, {Von Forstner}, {Baz{\'a}n}, {Akira Stern}, {Evans}, {Jain},
  {Malocha}, {Ghosh}, {Sta{\'n}czak}, {SophieLemos}, {Ranjan Singh}, {De
  Visscher}, {Verma}, {Airmansmith97}, {Buddhika}, {Sharma}, {Pathak},
  {Rideout}, {Agrawal}, {Alam}, {Bates}, {Park}, {Shukla}, {Mishra}, {Dubey},
  {Taylor}, {Dacie}, {Jacob}, {Goel}, {Sharma}, {Inchaurrandieta}, {Cetusic},
  {Reiter}, {Zahniy}, {Sidhu}, {Bray}, {Meszaros}, {Eigenbrot}, {Surve},
  {Parkhi}, {Robitaille}, {Pandey}, {Price-Whelan}, {J}, {Chicrala}, {Ankit},
  {Guennou}, {D'Avella}, {Williams}, {Verma}, {Ballew}, {Murphy}, {Lodha},
  {Bose}, {Augspurger}, {Krishan}, {Honey}, {Neerajkulk}, {Altunian}, {Ranjan},
  {Bhope}, {Molina}, {Gomillion}, {Kothari}, {Streicher}, {Wiedemann},
  {Mampaey}, {Nomiya}, {Mridulpandey}, {Habib}, {Letts}, {Agarwal}, {Singh
  Gaba}, {Hill}, {Ke{\c{s}}kek}, {Kumar}, {Verstringe}, {Mackenzie Dover},
  {Tollerud}, {Arias}, {Srikanth}, {Jain}, {Stone}, {Kustov}, {Smith}, {Sinha},
  {Kannojia}, {Mehrotra}, {Yadav}, {Paul}, {Wilkinson}, {Caswell}, {Braccia},
  {Pereira}, {Gates}, {Yasintoda}, {Kien Dang}, {Wilson}, {Bankar},
  {Bahuleyan}, {B}, {Platipo}, {Stevens}, {Gyenge}, {Schoentgen},
  {Shahdadpuri}, {Dedhia}, {Mendero}, {Cheung}, {Agrawal}, {Mangaonkar},
  {Lyes}, {Resakra}, {Ghosh}, {Hiware}, {Chaudhari}, {Reddy Mekala}, {Krishna},
  {Buitrago-Casas}, {Das}, {Mishra}, {Sharma}, {Wimbish}, {Calixto},
  {Babuschkin}, {Mathur}, {Murray}, \&
  {Nakul-Shahdadpuri}}]{sunpy2022zndo....591887M}
{Mumford}, S.~J., {Freij}, N., {Christe}, S., {et~al.} 2022, {SunPy}, v3.1.5,
  Zenodo,  Zenodo, \dodoi{10.5281/zenodo.591887}

\bibitem[{{Ni} {et~al.}(2010){Ni}, {Germaschewski}, {Huang}, {Sullivan},
  {Yang}, \& {Bhattacharjee}}]{Ni_2010PhPl...17e2109N}
{Ni}, L., {Germaschewski}, K., {Huang}, Y.-M., {et~al.} 2010, Physics of
  Plasmas, 17, 052109, \dodoi{10.1063/1.3428553}

\bibitem[{{O'Dwyer} {et~al.}(2010){O'Dwyer}, {Del Zanna}, {Mason}, {Weber}, \&
  {Tripathi}}]{O'Dwyer_etal_2010A&A...521A..21O}
{O'Dwyer}, B., {Del Zanna}, G., {Mason}, H.~E., {Weber}, M.~A., \& {Tripathi},
  D. 2010, \aap, 521, A21, \dodoi{10.1051/0004-6361/201014872}

\bibitem[{{Orlando} {et~al.}(2003){Orlando}, {Peres}, {Reale}, {Rosner}, \&
  {Siegel}}]{Orlando2003MmSAI..74..643O}
{Orlando}, S., {Peres}, G., {Reale}, F., {Rosner}, R., \& {Siegel}, A. 2003,
  \memsai, 74, 643

\bibitem[{{Parker}(1957)}]{Parker1957JGR....62..509P}
{Parker}, E.~N. 1957, \jgr, 62, 509, \dodoi{10.1029/JZ062i004p00509}

\bibitem[{{Petschek}(1964)}]{Petschek1964NASSP..50..425P}
{Petschek}, H.~E. 1964, in NASA Special Publication, Vol.~50, 425

\bibitem[{{Reeves} \& {Golub}(2011)}]{reeves2011}
{Reeves}, K.~K., \& {Golub}, L. 2011, ApJL, 727, L52,
  \dodoi{10.1088/2041-8205/727/2/L52}

\bibitem[{{Reeves} {et~al.}(2010){Reeves}, {Linker}, {Miki{\'c}}, \&
  {Forbes}}]{Reeves_2010ApJ...721.1547R}
{Reeves}, K.~K., {Linker}, J.~A., {Miki{\'c}}, Z., \& {Forbes}, T.~G. 2010,
  \apj, 721, 1547, \dodoi{10.1088/0004-637X/721/2/1547}

\bibitem[{{Reeves} {et~al.}(2019){Reeves}, {T{\"o}r{\"o}k}, {Miki{\'c}},
  {Linker}, \& {Murphy}}]{Reeves2019ApJ...887..103R}
{Reeves}, K.~K., {T{\"o}r{\"o}k}, T., {Miki{\'c}}, Z., {Linker}, J., \&
  {Murphy}, N.~A. 2019, \apj, 887, 103, \dodoi{10.3847/1538-4357/ab4ce8}

\bibitem[{{Rosner} {et~al.}(1986){Rosner}, {Low}, \&
  {Holzer}}]{Rosner1986psun....2..135R}
{Rosner}, R., {Low}, B.~C., \& {Holzer}, T.~E. 1986, in Physics of the Sun.
  Volume 2, Vol.~2, 135--180

\bibitem[{{Savage} {et~al.}(2010){Savage}, {McKenzie}, {Reeves}, {Forbes}, \&
  {Longcope}}]{savage2010}
{Savage}, S.~L., {McKenzie}, D.~E., {Reeves}, K.~K., {Forbes}, T.~G., \&
  {Longcope}, D.~W. 2010, ApJ, 722, 329, \dodoi{10.1088/0004-637X/722/1/329}

\bibitem[{{Schmelz} {et~al.}(2012){Schmelz}, {Reames}, {von Steiger}, \&
  {Basu}}]{Schmelz2012ApJ...755...33S}
{Schmelz}, J.~T., {Reames}, D.~V., {von Steiger}, R., \& {Basu}, S. 2012, \apj,
  755, 33, \dodoi{10.1088/0004-637X/755/1/33}

\bibitem[{{Seaton} {et~al.}(2017){Seaton}, {Bartz}, \&
  {Darnel}}]{Seaton2017ApJ...835..139S}
{Seaton}, D.~B., {Bartz}, A.~E., \& {Darnel}, J.~M. 2017, \apj, 835, 139,
  \dodoi{10.3847/1538-4357/835/2/139}

\bibitem[{{Seaton} \& {Forbes}(2009)}]{Seaton2009ApJ...701..348S}
{Seaton}, D.~B., \& {Forbes}, T.~G. 2009, \apj, 701, 348,
  \dodoi{10.1088/0004-637X/701/1/348}

\bibitem[{{Shen} {et~al.}(2018){Shen}, {Kong}, {Guo}, {Raymond}, \&
  {Chen}}]{Shen_2018ApJ...869..116S}
{Shen}, C., {Kong}, X., {Guo}, F., {Raymond}, J.~C., \& {Chen}, B. 2018, \apj,
  869, 116, \dodoi{10.3847/1538-4357/aaeed3}

\bibitem[{{Shen} {et~al.}(2011){Shen}, {Lin}, \& {Murphy}}]{shen2011}
{Shen}, C., {Lin}, J., \& {Murphy}, N.~A. 2011, \apj, 737, 14,
  \dodoi{10.1088/0004-637X/737/1/14}

\bibitem[{{Shen} {et~al.}(2013{\natexlab{a}}){Shen}, {Lin}, {Murphy}, \&
  {Raymond}}]{shen2013b}
{Shen}, C., {Lin}, J., {Murphy}, N.~A., \& {Raymond}, J.~C. 2013{\natexlab{a}},
  Physics of Plasmas, 20, 072114, \dodoi{10.1063/1.4816711}

\bibitem[{{Shen} {et~al.}(2017){Shen}, {Raymond}, {Miki{\'c}}, {Linker},
  {Reeves}, \& {Murphy}}]{Shen2017ApJ...850...26S}
{Shen}, C., {Raymond}, J.~C., {Miki{\'c}}, Z., {et~al.} 2017, \apj, 850, 26,
  \dodoi{10.3847/1538-4357/aa93f3}

\bibitem[{{Shen} {et~al.}(2015){Shen}, {Raymond}, \&
  {Murphy}}]{Shen2015zndo....272609S}
{Shen}, C., {Raymond}, J.~C., \& {Murphy}, N.~A. 2015, {Non-Equilibrium
  Ionization Code}, Zenodo,  Zenodo, \dodoi{10.5281/zenodo.272609}

\bibitem[{Shen {et~al.}(2015)Shen, Raymond, Murphy, \& Lin}]{Shen2015}
Shen, C., Raymond, J.~C., Murphy, N.~A., \& Lin, J. 2015, Astronomy and
  Computing, 12, 1, \dodoi{10.1016/j.ascom.2015.04.003}

\bibitem[{{Shen} {et~al.}(2013{\natexlab{b}}){Shen}, {Reeves}, {Raymond},
  {Murphy}, {Ko}, {Lin}, {Miki{\'c}}, \& {Linker}}]{shen2013a}
{Shen}, C., {Reeves}, K.~K., {Raymond}, J.~C., {et~al.} 2013{\natexlab{b}},
  ApJ, 773, 110, \dodoi{10.1088/0004-637X/773/2/110}

\bibitem[{{Shibata} \& {Magara}(2011)}]{Shibata2011LRSP....8....6S}
{Shibata}, K., \& {Magara}, T. 2011, Living Reviews in Solar Physics, 8, 6,
  \dodoi{10.12942/lrsp-2011-6}

\bibitem[{{Smith} \& {Hughes}(2010)}]{Smith2010ApJ...718..583S}
{Smith}, R.~K., \& {Hughes}, J.~P. 2010, \apj, 718, 583,
  \dodoi{10.1088/0004-637X/718/1/583}

\bibitem[{{Stone} {et~al.}(2008){Stone}, {Gardiner}, {Teuben}, {Hawley}, \&
  {Simon}}]{Stone2008ApJS..178..137S}
{Stone}, J.~M., {Gardiner}, T.~A., {Teuben}, P., {Hawley}, J.~F., \& {Simon},
  J.~B. 2008, \apjs, 178, 137, \dodoi{10.1086/588755}

\bibitem[{{Sturrock}(1968)}]{Sturrock1968}
{Sturrock}, P.~A. 1968, in IAU Symposium, Vol.~35, Structure and Development of
  Solar Active Regions, ed. K.~O. {Kiepenheuer}, 471

\bibitem[{{Sweet}(1958)}]{Sweet_1958IAUS....6..123S}
{Sweet}, P.~A. 1958, in IAU Symposium, Vol.~6, Electromagnetic Phenomena in
  Cosmical Physics, ed. B.~{Lehnert}, 123

\bibitem[{{Szenicer} {et~al.}(2019){Szenicer}, {Fouhey}, {Munoz-Jaramillo},
  {Wright}, {Thomas}, {Galvez}, {Jin}, \&
  {Cheung}}]{Szenicer2019SciA....5.6548S}
{Szenicer}, A., {Fouhey}, D.~F., {Munoz-Jaramillo}, A., {et~al.} 2019, Science
  Advances, 5, eaaw6548, \dodoi{10.1126/sciadv.aaw6548}

\bibitem[{{Warren} {et~al.}(2018){Warren}, {Brooks}, {Ugarte-Urra}, {Reep},
  {Crump}, \& {Doschek}}]{Warren2018ApJ...854..122W}
{Warren}, H.~P., {Brooks}, D.~H., {Ugarte-Urra}, I., {et~al.} 2018, \apj, 854,
  122, \dodoi{10.3847/1538-4357/aaa9b8}

\bibitem[{Winter {et~al.}(2011)Winter, Martens, \&
  Reeves}]{winter_simulating_2011}
Winter, H.~D., Martens, P., \& Reeves, K.~K. 2011, The Astrophysical Journal,
  735, 103, \dodoi{10.1088/0004-637X/735/2/103}

\bibitem[{{Yokoyama} \& {Shibata}(1997)}]{Yokoyama1997ApJ...474L..61Y}
{Yokoyama}, T., \& {Shibata}, K. 1997, \apjl, 474, L61, \dodoi{10.1086/310429}

\bibitem[{{Yokoyama} \& {Shibata}(2001)}]{Yokoyama2001ApJ...549.1160Y}
---. 2001, \apj, 549, 1160, \dodoi{10.1086/319440}

\bibitem[{{Zhang} {et~al.}(2018){Zhang}, {Foster}, \&
  {Smith}}]{Zhang_2018ApJ...864...79Z}
{Zhang}, G.-Y., {Foster}, A., \& {Smith}, R. 2018, \apj, 864, 79,
  \dodoi{10.3847/1538-4357/aad692}

\bibitem[{{Zhang} {et~al.}(2019){Zhang}, {Slavin}, {Foster}, {Smith}, {ZuHone},
  {Zhou}, \& {Chen}}]{Zhang2019ApJ...875...81Z}
{Zhang}, G.-Y., {Slavin}, J.~D., {Foster}, A., {et~al.} 2019, \apj, 875, 81,
  \dodoi{10.3847/1538-4357/ab0f9a}

\end{thebibliography}
\bibliographystyle{aasjournal}



\end{document}